\documentclass{article}
\usepackage{amssymb}
\usepackage{amsmath}

\input amssym.def
\input amssym

\newtheorem{theorem}{Theorem}
\newtheorem{lemma}{Lemma}

\newtheorem{ex}{Example}

\newcommand{\C}{\mathbb C} 
\newcommand{\R}{\mathbb R} 
\newcommand{\N}{\mathbb N} 

\newcommand{\mB}{\mathcal{B}} 
\newcommand{\mE}{\mathcal{E}}

\newcommand{\mK}{\mathcal{K}}
\newcommand{\mM}{\mathcal{M}}
\newcommand{\mN}{\mathcal{N}}
\newcommand{\mP}{\mathcal{P}}
\newcommand{\mS}{\mathcal{S}}
\newcommand{\mT}{\mathcal{T}}

\newcommand{\hi}{\mathcal{H}} 
\newcommand{\tsh}{\mathcal{T}_s(\hi)} 
\newcommand{\bsh}{\mathcal{B}_s(\hi)} 
\newcommand{\csh}{\mathcal{C}_s(\hi)} 
\newcommand{\sh}{\mathcal{S}(\hi)} 
\newcommand{\esh}{\partial_e\sh} 
\newcommand{\eh}{\mathcal{E}(\hi)} 
\newcommand{\ph}{\mathcal{P}(\hi)} 
\newcommand{\pk}{\mP(\mK)} 

\newcommand{\os}{(\Omega,\Sigma)} 
\newcommand{\phx}{(\ph,\Xi)} 

\newcommand{\mos}{\mathcal{M}_{\mathbb R}\os} 
\newcommand{\fos}{\mathcal{F}_{\mathbb R}\os} 
\newcommand{\sos}{\mathcal{S}\os} 
\newcommand{\eos}{\mathcal{E}\os} 
\newcommand{\mpx}{\mathcal{M}_{\mathbb R}\phx} 
\newcommand{\fpx}{\mathcal{F}_{\mathbb R}\phx} 
\newcommand{\spx}{\mathcal{S}\phx} 
\newcommand{\epx}{\mathcal{E}\phx} 

\newcommand{\sigtop}[1]{\sigma({#1})} 
\newcommand{\dual}[2]{\langle {#1},{#2}\rangle} 
\newcommand{\tr}[1]{\mathrm{tr} \, {#1}} 
\newcommand{\no}[1]{\left\|{#1}\right\|} 
\newcommand{\kb}[2]{|#1\,\rangle\langle\,#2|} 
\newcommand{\ip}[2]{\langle {#1}|{#2}\rangle} 
\newcommand{\delo}{\delta_{\omega}} 
\newcommand{\delp}{\delta_P} 

\newcommand{\set}[2]{\{{#1} \, | \, {#2}\}} 

\newcommand{\omti}{\widetilde{\Omega}}
\newcommand{\sigti}{\widetilde{\Sigma}}
\newcommand{\nuti}{\tilde{\nu}}
\newcommand{\muti}{\tilde{\mu}}
\newcommand{\deloti}{\tilde{\delta}_{\omega}} 
\newcommand{\rti}{\widetilde R} 
\newcommand{\mosti}{\mM_{\R}(\omti,\sigti)}
\newcommand{\fosti}{\mathcal{F}_{\R}(\omti,\sigti)}
\newcommand{\sosti}{\mathcal{S}(\omti,\sigti)}
\newcommand{\sn}{\mathcal{S}_{\mN}}

\newcommand{\fii}{\varphi}

\newenvironment{proof}[1][
Proof]{\textbf{#1.} }{\ \rule{0.5em}{0.5em}}
\newcommand{\qed}{\ $\square$}

\begin{document}

\title{The Structure of Classical Extensions of Quantum Probability
Theory\thanks{Dedicated to Paula and Stan Gudder}
 }

\author{Werner Stulpe\thanks{Electronic address: stulpe@fh-aachen.de}\\
{\small Aachen University of Applied Sciences, J\"ulich Campus,
D-52428, Germany}
\vspace{12pt}\\
Paul Busch\thanks{Electronic address: pb516@york.ac.uk}\\
{\small Perimeter Institute for Theoretical Physics, Waterloo, Canada}\\
{\small and Department of Mathematics, University of York, UK} }

\date{}
\maketitle

\begin{abstract}
\noindent On the basis of a suggestive definition of a classical
extension of quantum mechanics in terms of statistical models, we
prove that every such classical extension is essentially given by
the so-called Misra-Bugajski reduction map. We consider how this map
enables one to understand quantum mechanics as a reduced classical
statistical theory on the projective Hilbert space as phase space
and discuss features of the induced hidden-variables model.
Moreover, some relevant technical results on the
topology and Borel structure of the projective Hilbert space are reviewed.\\
\mbox{}
\newline
Running Title: Classical Extensions of Quantum Probability Theory
\end{abstract}

\section{Introduction}

Every statistical (probabilistic) physical theory can be based on a
set $ \mS $ of {\it states}, a set $ \mE $ of {\it effects}, and a
{\it probability functional} associating each state $ s \in \mS $
and each effect $ a \in \mE $ with a real number $ \dual{s}{a} \in
[0,1] $, the latter being the {\it probability for the outcome `yes'
of the effect $a$ in the state $s$}
\cite{lud70;85,lud83,dav70,dav76,gud79}. We summarize these basic
concepts of a statistical theory by the pair $ \dual{\mS}{\mE} $; we
call $ \dual{\mS}{\mE} $ a {\it statistical model} if the following
properties are satisfied \cite{hol82,hol01,bel95a;b,bel97}. Since
states can be mixed, $ \mS $ has to be closed under such mixtures,
and the probability functional must be affine  in the states
(mixture-preserving); moreover, we assume that the states and the
effects separate each other (i.e., $ \dual{s_1}{a} = \dual{s_2}{a} $
for all $ a \in \mE $ implies $ s_1 = s_2 $,
 and $ \dual{s}{a_1} = \dual{s}{a_2} $ for all $ s \in \mS $
implies $ a_1 = a_2 $).

Given a statistical model $ \dual{\mS_1}{\mE_1} $, assume only a subset
$ \mE_2 \subseteq \mE_1 $ is accessible. In general, $ \mE_2 $ no longer
separates $ \mS_1 $; call two states $ s,\tilde{s} \in \mS_1 $ equivalent if
$ \dual{s}{a} = \dual{\tilde{s}}{a} $ for all $ a \in \mE_2 $. Let $ \mS_2 $
be the set of the equivalence classes and define
\begin{equation}\label{dual1}
\dual{[s]}{a} := \dual{s}{a}
\end{equation}
where $ [s] \in \mS_2 $ and $ a \in \mE_2 $. Then $ \mS_2 $ is a new set
of states and $ \dual{\mS_2}{\mE_2} $ a new statistical model;
$ \dual{\mS_2}{\mE_2} $ is a {\it reduction of $ \dual{\mS_1}{\mE_1} $}, and
$ \dual{\mS_1}{\mE_1} $ is an {\it extension of
$ \dual{\mS_2}{\mE_2} $}. Let $ R \! : \mS_1 \to \mS_2 $ be
the canonical projection, i.e., $ R(s) := [s] $, and define
the embedding map $ R' \! : \mE_2 \to \mE_1 $, i.e.,
$ R'(a) := a $. Then Eq.~(\ref{dual1}) can be written as
\[
\dual{R(s)}{a} = \dual{s}{R'(a)}.
\]
Note that $R$ is affine and surjective, whereas $R'$ is injective. We call
$R$ a {\it reduction map}.

Next let $ \dual{\mS_1}{\mE_1} $ and $ \dual{\mS_2}{\mE_2} $ be two
arbitrary statistical models and $ R \! : \mS_1 \to \mS_2 $ a surjective
affine mapping. Observe that $ s_1 \mapsto \dual{R(s_1)}{a_2} $ is an affine
functional on $ \mS_1 $ with values in the interval $ [0,1] $; assume that,
for each effect $ a_2 \in \mE_2 $, there exists an effect
$ a_1 \in \mE_1 $ such that
\begin{equation}\label{dual2}
\dual{R(s_1)}{a_2} = \dual{s_1}{a_1}
\end{equation}
holds for all $ s_1 \in \mS_1 $. Clearly, $ a_1 $ is uniquely determined,
and we can define a map $ R' \! : \mE_2 \to \mE_1 $ according to
$ R'(a_2) := a_1 $. Then Eq.~(\ref{dual2}) reads
\begin{equation}\label{dual3}
\dual{R(s_1)}{a_2} = \dual{s_1}{R'(a_2)},
\end{equation}
and one easily shows that $R'$ is injective. Moreover, we can call
two states $ s_1,\tilde{s}_1 \in \mS_1 $ equivalent if $ R(s_1) =
R(\tilde{s}_1) $; for effects of the form $ R'(a_2) $, such
equivalent states $ s_1 $ and $ \tilde{s}_1 $ give rise to the same
probabilities. Because $R$ is surjective, the states $ s_2 \in S_2 $
can be identified with the equivalence classes $ [s_1] =
R^{-1}(\{s_2\}) $ where $ s_2 = Rs_1 $. Because $ R' $ is injective,
we can further identify the effects $ a_2 \in \mE_2 $ with the
effects $ R'(a_2) $, i.e., $ \mE_2 $ can be considered as a subset
of $ \mE_1 $. By means of these identifications, Eq.~(\ref{dual3})
coincides with Eq.~(\ref{dual1}), and $R$ takes the role of the
canonical projection. Hence, the relation between the two
statistical models of this paragraph is the same as that between the
two statistical models of the preceding paragraph.

If $ \dual{\mS_1}{\mE_1} $ and $ \dual{\mS_2}{\mE_2} $ are two
statistical models and $R$ is a surjective affine mapping from $
\mS_1 $ onto $ \mS_2 $ for which, in the sense just described, a
mapping $R'$ exists, then we call $ \dual{\mS_2}{\mE_2} $ a {\it
reduction of $ \dual{\mS_1}{\mE_1} $}, $ \dual{\mS_1}{\mE_1} $ an
{\it extension of $ \dual{\mS_2}{\mE_2} $}, and $R$ a {\it reduction
map}. Since statistical models can be embedded into dual pairs of
vector spaces (one vector space being a base-norm space and the
other one an order-unit norm space, the pair forming a so-called
{\em statistical duality} \cite{lud70;85,lud83,wer83,hol82}), the
reduction-extension concept for statistical models can be
reformulated in this general context. The reduction map $R$ is then
a surjective bounded linear map, and $R'$ is the adjoint map of $R$
which is linear, bounded, and injective. We do not consider this
reformulation in complete generality, instead we shall study a
reduction-extension concept specific to the subject of this paper
which concerns the relation between classical and quantum
probability.

It is the aim of this paper to revisit a particular classical
extension of quantum mechanics defined by what we call the {\em
Misra-Bugajski reduction map}
\cite{mis74,ghi76,hol82,bug91;93a-d,bel95a;b,stu01,bus04}, and to
show that this map is essentially the only possible reduction map
from a classical statistical model to the quantum statistical model,
i.e., essentially the only possible way to obtain a classical
extension of quantum probability theory. To this end, we first
define the notions of quantum and classical statistical model. In
doing so we also introduce most of the notations used in the paper.

Let a complex separable Hilbert space $ \hi \neq \{0\} $ be given. We denote
the real vector space of the self-adjoint trace-class operators by $ \tsh $
and the convex set of the positive trace-class operators of trace $1$ by
$ \sh $; the operators of $ \sh $ are the density operators and describe
the quantum states. The pair $ (\tsh,\sh) $ is a base-normed Banach space
with closed positive cone, the base norm being the trace norm. We denote
the real vector space of all bounded self-adjoint operators by $ \bsh $
and the unit operator by $I$. The pair $ (\bsh,I) $ where $ \bsh $ is
equipped with its order relation, is an order-unit normed Banach space
with closed positive cone, the norm being the usual operator norm. The
elements of the order-unit interval $ \eh := [0,I] $ describe the
quantum mechanical effects. As is well known, $ \bsh $ can be
considered as the dual space $ (\tsh)' $ where the duality is
given by the trace functional
\[
(V,A) \mapsto \dual{V}{A} := \tr{VA},
\]
$ V \in \tsh $, $ A \in \bsh $. The restriction of this bilinear
functional to $ \sh \times \eh $ is the quantum probability
functional; $ \tr{WA} $ is the probability for the outcome `yes' of
the effect $ A \in \eh $ in the state $ W \in \sh $. Thus, $
\dual{\tsh}{\bsh} $ is a dual pair of vector spaces (in fact a
statistical duality) and $ \dual{\sh}{\eh} $ the {\it quantum
statistical model} \cite{lud83,dav76,bus95,hol01}.

Further we recall that the extreme points of the convex set $ \sh $, i.e.,
the pure quantum states, are the one-dimensional orthogonal projections
$ P = P_{\fii}: = \kb{\fii}{\fii} $, $ \no{\fii} = 1 $. We denote the set
of these extreme points, i.e., the extreme boundary, by $ \esh $. The
extreme points of the convex set $ \eh $ are all orthogonal projections,
these are sometimes called {\it sharp} effects whereas the other ones
are called {\it unsharp} effects.---We also recall that
$ \sigtop{\tsh,\bsh} $ is the weak Banach-space topology of $ \tsh $,
i.e., the coarsest topology on $ \tsh $ in which the elements of $ \bsh $,
considered as linear functionals on $ \tsh $, are continuous.

For a general measurable space $ \os $ where $ \Omega $ is a nonempty set
and $ \Sigma $ an arbitrary $ \sigma $-algebra of subsets of $ \Omega $,
let $ \mos $ be the real vector space of the real-valued measures on
$ \os $ (i.e., of the $ \sigma $-additive real-valued set functions on
$ \Sigma $). We denote the convex subset of the positive normalized
measures by $ \sos $; the elements of $ \sos $ are probability measures
and describe classical states. The pair $ (\mos,\sos) $ is a base-normed
Banach space with closed positive cone, the base norm being the
total-variation norm. By $ \fos $ we denote the real vector space
of the bounded $ \Sigma $-measurable functions on $ \Omega $ and by
$ \chi_E $ the characteristic function of a set $ E \in \Sigma $. The pair
$ (\fos,\chi_{\Omega}) $ together with the order relation of $ \fos $ is
an order-unit normed Banach space with closed positive cone, the order-unit
norm being the supremum norm. The elements of the order-unit interval
$ \eos := [0,\chi_{\Omega}] $ describe the classical effects. By the
bilinear functional given by the integral
\[
(\nu,f) \mapsto \dual{\nu}{f} := \int_{\Omega} fd\nu,
\]
$ \nu \in \mos $, $ f \in \fos $, the spaces $ \mos $ and $ \fos $
are placed in duality to each other; in particular, $ \fos $ can be
considered as a norm-closed subspace of the dual space $ (\mos)' $
where in general the dual space is larger than $ \fos $. The
restriction of $ (\nu,f) \mapsto \dual{\nu}{f} $ to $ \sos \times
\eos $ is the classical probability functional; $ \int fd\nu $ is
the probability for the outcome `yes' of the effect $ f \in \eos $
in the state $ \mu \in \sos $. Again, $ \dual{\mos}{\fos} $ is a
dual pair of vector spaces (a statistical duality), whereas $
\dual{\sos}{\eos} $ is the {\it classical statistical model}
\cite{dav70,gud79,stu86,sin92,bug96,bug98,gud98}.

We remark that the Dirac measures $ \delo $, $ \omega \in \Omega $,
are extreme points of the convex set $ \sos $, but in general there
are also other extreme points. The extreme points of the convex set
$ \eos $ are the characteristic functions $ \chi_E $, $ E \in \Sigma
$, these are the {\it sharp} classical effects (in the terminology
of classical probability theory, the {\it events}), the other
effects are {\it unsharp} or {\it fuzzy}.---Finally, we recall that
$ \sigtop{\mos,\fos} $ is the coarsest topology on $ \mos $ in which
the elements of $ \fos $, considered as linear functionals on $ \mos
$, are continuous.

Now assume that, for the two statistical models
$ \dual{\mS_1}{\mE_1} = \dual{\sos}{\eos} $ and
$ \dual{\mS_2}{\mE_2} = \dual{\sh}{\eh} $, a reduction map
$ R \! : \sos \to \sh $ is given. It is not hard to show that
the surjective affine mapping $R$ can uniquely be extended to
a surjective linear map from $ \mos $ onto $ \tsh $ which
we also call $R$; the linear map $R$ is
automatically positive and bounded. According to Eq.~(\ref{dual3}) the
injective mapping $ R' \! : \eh \to \eos $ satisfies
\begin{equation}\label{dual4}
\tr{(R\mu)A} = \dual{R\mu}{A} = \dual{\mu}{R'A} = \int_{\Omega} R'A \, d\mu
\end{equation}
for all $ \mu \in \sos $ and all $ A \in \eh $; $R'$ is also
affine. Moreover, from (\ref{dual4}) it follows that the adjoint map of
$R$ w.r.t.\ the dual pairs $ \dual{\mos}{\fos} $ and $ \dual{\tsh}{\bsh} $
exists, this adjoint map $ R' \! : \bsh \to \fos $ is a unique linear
extension of the affine mapping $ R' \! : \eh \to \eos $ and is also
injective.

The existence of the adjoint map $R'$ w.r.t.\ the considered dual
pairs is equivalent to $ R^*\bsh \subseteq \fos $ where $ R^* \! :
\bsh \to (\mos)' $ is the Banach-space adjoint map of $R$. According
to general results in duality theory, the existence of the linear
map $R'$ is also equivalent to the $ \sigtop{\mos,\fos} $-$
\sigtop{\tsh,\bsh} $ continuity of $R$.---The crucial properties of
the linear map $R$ are summarized in the following definition.
\vspace{9pt}

\noindent {\bf Definition} \ We call a linear map $ R \! : \mos
\to \tsh $ a {\it reduction map} if
\begin{enumerate}
\item[(i)] $ R\sos = \sh $;
\item[(ii)] $R$ is $ \sigtop{\mos,\fos}$-$\sigtop{\tsh,\bsh} $-continuous.
\end{enumerate}
We will say that the linear map $R$ (or its affine restriction)
together with the dual map $R'$ constitutes a {\it reduction} of the
classical statistical model $ \dual{\sos}{\eos} $ to the quantum
statistical model $ \dual{\sh}{\eh} $. In particular, we will say
that $R$ and $R'$ constitute a {\it classical extension of quantum
mechanics}.
\vspace{9pt}

\indent The properties of $R$ stated in this definition
imply again that $R$ is bounded, positive, and surjective and that
$R'$ exists and is injective. Furthermore, one easily shows that
$R'$ is positive and that $ R'I = \chi_{\Omega} $ and $ R'\eh
\subseteq \eos $. The restrictions of $R$ and $R'$ to $ \sos $ and $
\eos $, respectively, are affine; clearly, the restriction of $R$ to
$ \sos $ is a reduction map as defined previously in the context of
two general statistical models $ \dual{\mS_1}{\mE_1} $ and $
\dual{\mS_2}{\mE_2} $.

It is not clear that classical extensions of quantum mechanics do
exist, in fact, this may be considered surprising. The typical
example of a reduction map is the so-called {\it Misra-Bugajski map}
which we present in Section~\ref{sec:mb}. In Section~\ref{sec:cextq}
we prove our result that every reduction map giving a classical
extension of quantum mechanics is essentially equivalent to the
Misra-Bugajski map. Thus, the Misra-Bugajski map is essentially
unique and yields a canonical classical extension of quantum
mechanics.

Sections \ref{sec:top} and \ref{sec:meas} provide prerequisite
results on the topology and the Borel structure of the projective
Hilbert space which will be identified with the extreme boundary $
\esh $ of $ \sh $. In Section \ref{sec:ex} some examples of
reduction maps different from the Misra-Bugajski map are presented.
Finally, in Section \ref{sec:int} the physical interpretation of the
results of Sections \ref{sec:mb} and \ref{sec:cextq} is discussed.

\section{The Topology of the Projective Hilbert Space}\label{sec:top}

In this section we undertake a systematic review and comparison,
sketched out in this context previously by Bugajski \cite{bug94}, of
the various topologies on the set of the pure quantum states or,
alternatively, on the projective Hilbert space associated with a
nontrivial separable complex Hilbert space $ \hi \neq \{0\} $.

Call two vectors of $ \hi^* := \hi \setminus \{0\} $ equivalent if
they differ by a complex factor, and define the {\it projective
Hilbert space $ \ph $} to be the set of the corresponding
equivalence classes which are often called {\it rays}. Instead of $
\hi^* $ one can consider only the unit sphere of $ \hi $, $ S := \{
\fii \in \hi \, | \, \no{\fii} = 1 \} $. Then two unit vectors are
called equivalent if they differ by a phase factor, and the set of
the corresponding equivalence classes, i.e., the set of the {\it
unit rays}, is denoted by $ S/S^1 $ (in this context, $ S^1 $ is
understood as the set of all phase factors, i.e., as the set of all
complex numbers of modulus $1$). Clearly, $ S/S^1 $ can be
identified with the projective Hilbert space $ \ph $. Furthermore,
we can consider the elements of $ \ph $ also as the one-dimensional
subspaces of $ \hi $ or, equivalently, as the one-dimensional
orthogonal projections $ P = P_{\fii} = \kb{\fii}{\fii} $, $
\no{\fii} = 1 $.

The set $ \hi^* $ and the unit sphere $S$ carry the topologies induced by
the metric topology of $ \hi $. Using the canonical projections
$ \mu \! : \hi^* \to \ph $, $ \mu(\fii) := [\fii] $, and
$ \nu \! : S \to S/S^1 $, $ \nu(\chi) := [\chi]_S $, where
$ [\fii] $ is a ray and $ [\chi]_S $ a unit ray, we can equip
the quotient sets $ \ph $ and $ S/S^1 $ with their quotient topologies
$ \mT_{\mu} $ and $ \mT_{\nu} $. Considering $ \mT_{\nu} $, a set
$ O \subseteq S/S^1 $ is called open if $ \nu^{-1}(O) $ is open.

\begin{theorem}\label{thm:ss1}
The set $ S/S^1 $, equipped with the quotient topology $ \mT_{\nu} $,
is a second-countable Hausdorff space, and $ \nu $ is an open continuous
mapping.
\end{theorem}
\proof{
By definition of $ \mT_{\nu} $, $ \nu $ is continuous. To show that
$ \nu $ is open, let $U$ be an open set of $S$. From
\[
\nu^{-1}(\nu(U)) = \nu^{-1}(\{ [\chi]_S \, | \, \chi \in U \})
                 = \bigcup_{\lambda \in S^1} \lambda U ,
\]
$ S^1 = \{ \lambda \in \C \, | \, |\lambda| = 1 \} $, it follows that
$ \nu^{-1}(\nu(U)) \subseteq S $ is open. So $ \nu(U) \subseteq S/S^1 $
is open; hence, $ \nu $ is open.

Next consider two different unit rays $ [\fii]_S $ and $ [\psi]_S $ where
$ \fii,\psi \in S $ and $ |\ip{\fii}{\psi}| = 1 - \varepsilon $,
$ 0 < \varepsilon \leq 1 $. Since the mapping
$ \chi \mapsto |\ip{\fii}{\chi}| $, $ \chi \in S $, is continuous,
the sets
\begin{equation}\label{u1}
U_1 := \left\{ \chi \in S \left| \,
              |\ip{\fii}{\chi}| > 1 - \tfrac{\varepsilon}{2} \right.
                                                             \right\}
\end{equation}
and
\begin{equation}\label{u2}
U_2 := \left\{ \chi \in S \left| \,
              |\ip{\fii}{\chi}| < 1 - \tfrac{\varepsilon}{2} \right.
                                                             \right\}
\end{equation}
are open neighborhoods of $ \fii $ and $ \psi $, respectively. Consequently,
the sets $ O_1 := \nu(U_1) $ and $ O_2 := \nu(U_2) $ are open
neighborhoods of $ [\fii]_S $ and $ [\psi]_S $, respectively. Assume
$ O_1 \cap O_2 \neq \emptyset $. Let $ [\xi]_S \in O_1 \cap O_2 $, then
$ [\xi]_S = \nu(\chi_1) = \nu(\chi_2) $ where $ \chi_1 \in U_1 $ and
$ \chi_2 \in U_2 $. It follows that $ \chi_1 $ and $ \chi_2 $ are equivalent,
so $ |\ip{\fii}{\chi_1}| = |\ip{\fii}{\chi_2}| $, in contradiction to
$ \chi_1 \in U_1 $ and $ \chi_2 \in U_2 $. Hence, $ O_1 $ and $ O_2 $
are disjoint, and $ \mT_{\nu} $ is separating.

Finally, let $ \mB = \{ U_n \, | \, n \in \N \} $ be a countable base of
the topology of $S$ and define the open sets $ O_n := \nu(U_n) $. We show
that $ \{ O_n \, | \, n \in \N \} $ is a base of $ \mT_{\nu} $. For
$ O \in \mT_{\nu} $, we have that $ \nu^{-1}(O) $ is an open set of
$S$ and consequently $ \nu^{-1}(O) = \bigcup_{n \in M} U_n $ where
$ U_n \in \mB $ and $ M \subseteq \N $. Since $ \nu $ is surjective,
it follows that
\[
O = \nu(\nu^{-1}(O)) = \nu \left( \bigcup_{n \in M} U_n \right)
                     = \bigcup_{n \in M} \nu(U_n)
                     = \bigcup_{n \in M} O_n .
\]
Hence, $ \{ O_n \, | \, n \in \N \} $ is a countable base of
$ \mT_{\nu} $. \qed}

Analogously, it can be proved that the topology $ \mT_{\mu} $ on $ \ph $
is separating and second-countable and that the canonical projection $ \mu $
is open (and continuous by the definition of $ \mT_{\mu} $). Moreover,
one can show that the natural bijection $ \beta \! : \ph \to S/S^1 $,
$ \beta([\fii]) := \left[ \frac{\fii}{\no{\fii}} \right]_S $,
$ \beta^{-1}([\chi]_S) = [\chi] $, is a homeomorphism. Thus, identifying
$ \ph $ and $ S/S^1 $ by $ \beta $, the topologies $ \mT_{\mu} $ and
$ \mT_{\nu} $ are the same.

The above definition of $ \ph $ and $ S/S^1 $ as well as of their
quotient topologies is related to a geometrical point of view. From
an operator-theoretical point of view, it is more obvious to identify
$ \ph $ with $ \esh $, the extreme boundary of $ \sh $, and to restrict
one of the various operator topologies to $ \esh $. A further definition
of a topology on $ \esh $ is suggested by the interpretation
of the one-dimensional projections $ P \in \esh $ as
the pure quantum states and by the requirement that
the transition probabilities between two pure states are
continuous functions. Next we consider, taking account of
$ \esh \subseteq \sh \subset \tsh \subseteq \bsh $, the metric topologies on
$ \esh $ induced by the trace-norm topology of $ \tsh $, resp.,
by the norm toplogy of $ \bsh$. After that we introduce the weak topology on
$ \esh $ defined by the transition-probability functions as well as
the restrictions of several weak operator topologies to $ \esh $. Finally,
we shall prove the surprising result that all the many toplogies on
$ \ph \cong S/S^1 \cong \esh $ are equivalent.

\begin{theorem}\label{thm:esh-dist}
Let $ P_{\fii} = \kb{\fii}{\fii} \in \esh $ and
$ P_{\psi} = \kb{\psi}{\psi} \in \esh $ where
$ \no{\fii} = \no{\psi} = 1 $. Then
\begin{enumerate}
\item[(a)]
\[ \rho_n(P_{\fii},P_{\psi}) := \no{P_{\fii} - P_{\psi}}
                              = \sqrt{1 - |\ip{\fii}{\psi}|^2}
                              = \sqrt{1 - \tr{P_{\fii} P_{\psi}}}
\]
where the norm $ \no{\cdot} $ is the usual operator norm;
\item[(b)]
\[ \rho_{\mathrm{tr}}(P_{\fii},P_{\psi})
               :=  \no{P_{\fii} - P_{\psi}}_{\mathrm{tr}}
                = 2\no{P_{\fii} - P_{\psi}},
\]
in particular, the metrics $ \rho_n $ and $ \rho_{\mathrm{tr}} $ on
$ \esh $ induced by the operator norm $ \no{\cdot} $ and the trace
norm $ \no{\cdot}_{\mathrm{tr}} $ are equivalent;
\item[(c)]
\[
\no{P_{\fii} - P_{\psi}} \leq \no{\fii - \psi},
\]
in particular, the mapping $ \fii \mapsto P_{\fii} $ from $S$ into
$ \esh $ is continuous, $ \esh $ being equipped with $ \rho_n $ or
$ \rho_{\mathrm{tr}} $.
\end{enumerate}
\end{theorem}
\proof{
To prove (a) and (b), assume $ P_{\fii} \neq P_{\psi} $, otherwise
the statements are trivial. Then the range of $ P_{\fii} - P_{\psi} $
is a two-dimensional subspace of $ \hi $ and is spanned by the two linearly
independent unit vectors $ \fii $ and $ \psi $. Since eigenvectors of
$P_{\fii} - P_{\psi} $ belonging to eigenvalues $ \lambda \neq 0 $
must lie in the range of $ P_{\fii} - P_{\psi} $, they can be written as
$ \chi = \alpha \fii + \beta \psi $. Therefore, the eigenvalue problem
$ (P_{\fii} - P_{\psi}) \chi = \lambda \chi $, $ \chi \neq 0 $,
is equivalent to the two linear equations
\begin{eqnarray*}
    (1- \lambda) \alpha + \ip{\fii}{\psi} \beta & = & 0   \\
-\ip{\psi}{\fii} \alpha -   (1 + \lambda) \beta & = & 0
\end{eqnarray*}
where $ \alpha \neq 0 $ or $ \beta \neq 0 $. It follows that
$ \lambda = \pm \sqrt{1 - |\ip{\fii}{\psi}|^2} =: \lambda_{1,2} $. Hence,
$ P_{\fii} - P_{\psi} $ has the eigenvalues
$ \lambda_1 $, $0$, and $ \lambda_2 $. Now, from
$ \no{P_{\fii} - P_{\psi}} = \max \{ |\lambda_1|,|\lambda_2| \} $ and
$ \no{P_{\fii} - P_{\psi}}_{\mathrm{tr}} = |\lambda_1| + |\lambda_2| $,
we obtain the statements (a) and (b).---From
\begin{eqnarray*}
\no{P_{\fii} - P_{\psi}}^2
&  =   & 1 - |\ip{\fii}{\psi}|^2
   =     \no{\fii - \ip{\psi}{\fii} \psi}^2
   =     \no{(I - P_{\psi}) \fii}^2                                  \\
& \leq & \no{(I - P_{\psi}) \fii}^2 + \no{\psi - P_{\psi} \fii}^2    \\
&  =   & \no{(I - P_{\psi}) \fii - (\psi - P_{\psi} \fii)}^2         \\
&  =   & \no{\fii - \psi}^2
\end{eqnarray*}
we conclude statement (c). \qed}

According to statement (b) of Theorem \ref{thm:esh-dist}, the metrics $ \rho_n $ and
$ \rho_{\mathrm{tr}} $ give rise to the same topology
$ {\mT}_n = {\mT}_{\mathrm{tr}} $ as well as to the same uniform structures.

\begin{theorem}\label{thm:esh-sepc}
Equipped with either of the two metrics $ \rho_n $ and
$ \rho_{\mathrm{tr}} $, $ \esh $ is separable and complete.
\end{theorem}
\proof{
As a metric subspace of the separable Hilbert space $ \hi $, the unit sphere
$S$ is separable. Therefore, by statement (c) of Theorem \ref{thm:esh-dist}, the metric space
$ (\esh,\rho_n) $ is separable and so is $ (\esh,\rho_{\mathrm{tr}}) $
(the latter, moreover, implies the trace-norm separability of
$ \tsh $). Now let $ \{ P_n \}_{n \in \N} $ be a Cauchy sequence in
$ (\esh,\rho_{\mathrm{tr}}) $. Then there exists an operator
$ A \in \tsh $ such that $ \no{P_n - A}_{\mathrm{tr}} \to 0 $ as well as
$\no{P_n - A} \to 0 $ as $ n \to \infty $ (remember that, on $ \tsh $,
$ \no{\cdot}_{\mathrm{tr}} $ is stronger than $ \no{\cdot} $). From
\begin{eqnarray*}
\no{P_n - A^2}    =     \no{A^2 - P_n^2}
               & \leq & \no{A^2 - AP_n} + \no{AP_n - P_n^2}    \\
               & \leq & \no{A} \no{A - P_n} + \no{A - P_n}     \\
               & \to  &  0
\end{eqnarray*}
as $ n \to \infty $ we obtain $ A = \lim_{n \to \infty} P_n = A^2 $;
moreover,
\[
\tr{A} = \tr{AI} = \lim_{n \to \infty} \tr{P_nI} = 1.
\]
Hence, $A$ is a one-dimensional orthogonal projection, i.e.,
$ A \in \esh $. \qed}

Next we equip $ \esh $ with the topology $ \mT_0 $ generated by
the functions
\begin{equation}\label{p}
P \mapsto h_Q(P) := \tr{PQ} = |\ip{\fii}{\psi}|^2
\end{equation}
where $ P = \kb{\psi}{\psi} \in \esh $, $ Q = \kb{\fii}{\fii} \in \esh $,
and $ \no{\psi} = \no{\fii} = 1 $. That is, $ \mT_0 $ is the coarsest
topology on $ \esh $ such that all the real-valued functions $ h_Q $
are continuous. Note that $ \tr{PQ} = |\ip{\fii}{\psi}|^2 $ can be
interpreted as the transition probability between the two pure states
$P$ and $Q$.

\begin{lemma}\label{lem:esh}
The set $ \esh $, equipped with the topology $ \mT_0 $, is
a second-countable Hausdorff space. A countable base of $ \mT_0 $
is given by the finite intersections of the open sets
\begin{equation}\label{uklm}
\begin{array}{crl}
U_{klm} & := & h_{Q_k}^{-1} \left( \,
                            \left] q_l - \frac{1}{m},q_l + \frac{1}{m}
                                         \right[ \, \right)   \vspace{2mm}\\
        &  = & \left\{ P \in \esh \left| \,
               \left| \tr{PQ_k} - q_l \right| < \frac{1}{m}
                                                \right. \right\}
\end{array}
\end{equation}
where $ \{ Q_k \}_{k \in \N} $ is a sequence of one-dimensional orthogonal
projections being $ \rho_n $-dense in $ \esh $, $ \{ q_l \}_{l \in \N} $
is a sequence of numbers being dense in $ [0,1] \subseteq \R $, and
$ m \in \N $.
\end{lemma}
\proof{
Let $ P_1 $ and $ P_2 $ be any two different one-dimensional
projections. Choosing $ Q = P_1 $ in (\ref{p}), we obtain
$ h_{P_1}(P_1) = 1 \neq h_{P_1}(P_2) = 1 - \varepsilon $,
$ 0 < \varepsilon \leq 1 $. The sets
\[
U_1 := \left\{ P \in \esh \left| \,
                h_{P_1}(P) > 1 - \tfrac{\varepsilon}{2} \right. \right\}
\]
and
\[
U_2 := \left\{ P \in \esh \left| \,
                h_{P_1}(P) < 1 - \tfrac{\varepsilon}{2} \right. \right\}
\]
(cf.\ Eqs.~(\ref{u1}) and (\ref{u2})) are disjoint open neighborhoods of $ P_1 $ and
$ P_2 $, respectively. So $ \mT_0 $ is separating.

For an open set $ O \subseteq \R $, $ h_Q^{-1}(O) $ is $ \mT_0 $-open. We
next prove that
\begin{equation}\label{u}
U := h_Q^{-1}(O) = \bigcup_{U_{klm} \subseteq U} U_{klm}
\end{equation}
with $ U_{klm} $ according to (\ref{uklm}). Let $ P \in U $. Then
there exists an $ \varepsilon > 0 $ such that the interval
$ ]h_Q(P) - \varepsilon,h_Q(P) + \varepsilon[ $ is contained in $O$. Choose
$ m_0 \in \N $ such that $ \frac{1}{m_0} < \frac{\varepsilon}{2} $,
and choose a member $ q_{l_0} $ of the sequence $ \{ q_l \}_{l \in \N} $
and a member $ Q_{k_0} $ of $ \{ Q_k \}_{k \in \N} $ such that
$ |\tr{PQ} - q_{l_0}| < \frac{1}{2m_0} $ and
$ \no{Q_{k_0} - Q} < \frac{1}{2m_0} $. It follows that
\begin{eqnarray*}
|\tr{PQ_{k_0}} - q_{l_0}|
& \leq & |\tr{PQ_{k_0}} - \tr{PQ}| + |\tr{PQ} - q_{l_0}|    \\
& \leq &  \no{Q_{k_0} - Q} + |\tr{PQ} - q_{l_0}|            \\
&  <   &  \tfrac{1}{m_0}
\end{eqnarray*}
which, by (\ref{uklm}), means that $ P \in U_{k_0l_0m_0} $. We further have
to show that $ U_{k_0l_0m_0} \subseteq U $. To that end, let
$ \widetilde{P} \in U_{k_0l_0m_0} $. Then, from
\[
   \bigl| \tr{\widetilde{P}Q} - \tr{PQ} \bigr| \leq
   \bigl| \tr{\widetilde{P}Q} - \tr{\widetilde{P}Q_{k_0}} \bigr|
 + \bigl| \tr{\widetilde{P}Q_{k_0}} - q_{l_0} \bigr| + |q_{l_0} - \tr{PQ}|
\]
where the first term on the right-hand side is again smaller than
$ \no{Q - Q_{k_0}} $ and, by (\ref{uklm}), the second term is smaller than
$ \frac{1}{m_0} $, it follows that
\[
\bigl| h_Q(\widetilde{P}) - h_Q(P) \bigr|
 =   \bigl| \tr{\widetilde{P}Q} - \tr{PQ} \bigr|
\leq \tfrac{1}{2m_0} + \tfrac{1}{m_0} + \tfrac{1}{2m_0}
 = \tfrac{2}{m_0}
 < \varepsilon.
\]
This implies that $ h_Q(\widetilde{P}) \in
{]h_Q(P) - \varepsilon,h_Q(P) + \varepsilon[} \subseteq O $, i.e.,
$ \widetilde{P} \in h_Q^{-1}(O) = U $. Hence,
$ U_{k_0l_0m_0} \subseteq U $.

Summarizing, we have shown that, for $ P \in U $,
$ P \in U_{k_0l_0m_0} \subseteq U $. Hence,
$ U \subseteq \bigcup_{U_{klm} \subseteq U} U_{klm} \subseteq U $,
and assertion (\ref{u}) has been proved. The finite intersections of sets
of the form $ U = h_Q^{-1}(O) $ constitute a basis of the topology
$ \mT_0 $. Since every set $ U = h_Q^{-1}(O) $ is the union of sets
$ U_{klm} $, the intersections of finitely many sets
$ U = h_Q^{-1}(O) $ is the union of finite intersections of sets
$ U_{klm} $. Thus, the finite intersections of the sets
$ U_{klm} $ constitute a countable base of $ \mT_0 $. \qed}

Later we shall see that the topological space $ (\esh,\mT_0) $ is
homeomorphic to $ (\esh,\mT_n) $ as well as to $ (S/S^1,\mT_{\nu})
$. So it is also clear by Theorem \ref{thm:esh-sepc} or Theorem
\ref{thm:ss1} that $ (\esh,\mT_0) $ is a second-countable Hausdorff
space. The reason for stating Lemma \ref{lem:esh} is that later we
shall make explicit use of the particular countable base given
there.

The weak operator topology on the space $ \bsh $ of the bounded
self-adjoint operators on $ \hi $ is the coarsest topology such that
the linear functionals
\[
A \mapsto \ip{\fii}{A\psi}
\]
where $ A \in \bsh $ and $ \fii,\psi \in \hi $, are continuous. It is
sufficient to consider only the functionals
\begin{equation}\label{afun}
A \mapsto \ip{\fii}{A\fii}
\end{equation}
where $ \fii \in \hi $ and $ \no{\fii} = 1 $. The topology $ \mT_w $
induced on $ \esh \subset \bsh $ by the weak operator topology is the
coarsest topology on $ \esh $ such that the restrictions of the linear
functionals (\ref{afun}) to $ \esh $ are continuous. Since these restrictions
are given by
\[
P \mapsto \ip{\fii}{P\fii} = \tr{PQ} = h_Q(P)
\]
where $ P \in \esh $ and $ Q := \kb{\fii}{\fii} \in \esh $, the topology
$ \mT_w $ on $ \esh $ is, according to (\ref{p}), just our topology $ \mT_0 $.

Now we compare the weak topology $ \mT_0 $ with the metric topology
$ \mT_n $.

\begin{theorem}\label{thm:top-esh}
The weak topology $ \mT_0 $ on $ \esh $ and the metric topology $ \mT_n $ on
$ \esh $ are equal.
\end{theorem}
\proof{
According to (\ref{p}), a neighborhood base of $ P \in \esh $ w.r.t.\ $ \mT_0 $
is given by the open sets
\begin{equation}\label{up}
\begin{array}{c}
U(P;Q_1,\ldots,Q_n;\varepsilon) \hspace{7.5cm}                \vspace{2mm}\\
\hspace{0.8cm}
\begin{array}{crl}
& := & \displaystyle{\bigcap_{i=1}^n h_{Q_i}^{-1}
        ( \, ]h_{Q_i}(P) - \varepsilon,h_{Q_i}(P) + \varepsilon[ \, )}
                                                              \vspace{2mm}\\
&  = & \bigl\{ \widetilde{P} \in \esh \, \bigl| \, \bigl|
        h_{Q_i}(\widetilde{P}) - h_{Q_i}(P) \bigr| < \varepsilon \
        {\rm for} \ i=1,\ldots,n \bigr\}                      \vspace{2mm}\\
&  = & \bigl\{ \widetilde{P} \in \esh \, \bigl| \, \bigl|
        \tr{\widetilde{P}Q_i} - \tr{PQ_i} \bigr| < \varepsilon   \
        {\rm for} \ i=1,\ldots,n \bigr\}
\end{array}
\end{array}
\end{equation}
where $ Q_1,\dots,Q_n \in \esh $ and $ \varepsilon > 0 $;
a neighborhood base of $P$ w.r.t.\ $ \mT_n $ is given by the open balls
\begin{equation}\label{kep}
K_{\varepsilon}(P) := \bigl\{ \widetilde{P} \in \esh \, \bigl| \,
                      \bigl\| \widetilde{P} - P \bigr\| < \varepsilon
                                                        \bigr\}.
\end{equation}
If $ \bigl\| \widetilde{P} - P \bigl\| < \varepsilon $, then
\[
\bigl| \tr{\widetilde{P}Q_i} - \tr{PQ_i} \bigr|
  =   \bigl| \tr{Q_i(\widetilde{P} - P)} \bigr|
 \leq \no{Q_i}_{\mathrm{tr}} \bigl\| \widetilde{P} - P \bigr\|
  =   \bigl\| \widetilde{P} - P \bigr\|
  <   \varepsilon;
\]
hence, $ K_{\varepsilon}(P) \subseteq U(P;Q_1,\ldots,Q_n;\varepsilon) $. To
show some converse inclusion, take account of Theorem \ref{thm:esh-dist}, part (a), and note
that
\[
\bigl\| \widetilde{P} - P \bigr\|^2
  = 1 - \tr{\widetilde{P}P}
  = \bigl| \tr{\widetilde{P}P} - \tr{PP} \bigr|.
\]
In consequence, by (\ref{up}) and (\ref{kep}),
$ U(P;P;\varepsilon^2) = K_{\varepsilon}(P) $. Hence,
$ \mT_0 = \mT_n $. \qed}

It looks surprising that the topolgies $ \mT_0 $ and
$ \mT_n $ coincide. In fact, consider the sequence
$ \{ P_{\fii_n} \}_{n \in \N} $ where the vectors
$ \fii_n \in \hi $ constitute an orthonormal system. Then,
w.r.t.\ the weak operator topology, $ P_{\fii_n} \to 0 $ as
$ n \to \infty $ whereas $ \no{P_{\fii_n} - P_{\fii_{n+1}}} =1 $
for all $ n \in \N $. However, $ 0 \not\in \esh $; so
$ \{ P_{\fii_n} \}_{n \in \N} $ is convergent neither w.r.t.\
$ \mT_w = \mT_0 $ nor w.r.t.\ $ \mT_n $. Finally, like in the case
of the weak operator topology, there is a natural uniform structure
inducing $ \mT_0 $. The uniform structures that are canonically related to
$ \mT_0 $ and $ \mT_n $ are different: $ \{ P_{\fii_n} \}_{n \in \N} $ is
a Cauchy sequence w.r.t.\ the uniform structure belonging to $ \mT_0 $
but not w.r.t.\ that belonging to $ \mT_n $, i.e., w.r.t.\ the metric
$ \rho_n $.

We remark that besides $ \mT_0 $ and $ \mT_w $ several further
weak topologies can be defined on $ \esh $. Let $ \csh $ be the Banach space of the compact
self-adjoint operators and remember that $ (\csh)' = \tsh $. So the weak
Banach-space topologies of $ \csh $, $ \tsh $, and $ \bsh $ as well as
the weak-* Banach-space topologies of $ \tsh $ and $ \bsh $ can
be restricted to $ \esh $, thus giving the topologies
$ \mT_1 := \sigma(\csh,\tsh) \cap \esh $,
$ \mT_2 := \sigma(\tsh,\csh) \cap \esh $,
$ \mT_3 := \sigma(\tsh,\bsh) \cap \esh $,
$ \mT_4 := \sigma(\bsh,\tsh) \cap \esh $, and
$ \mT_5 := \sigma(\bsh,(\bsh)') \cap \esh $. Moreover,
the strong operator topology induces a topology $ \mT_s $ on
$ \esh $. From the obvious inclusions
\[
\mT_w \subseteq \mT_1 \subseteq \mT_2 \subseteq \mT_3
      \subseteq \mT_{\mathrm{tr}} ,
\]
\[
\mT_1 = \mT_4 \subseteq \mT_5 = \mT_1 ,
\]
and
\[
\mT_w \subseteq \mT_s \subseteq \mT_n
\]
as well as from the shown equality
\[
\mT_0 = \mT_w = \mT_n = \mT_{\mathrm{tr}}
\]
it follows that the topologies $ \mT_1,\ldots,\mT_5 $ and $ \mT_s $
also coincide with $ \mT_0 $.

Finally, we show that all the topologies on $ \esh $ are equivalent to
the quotient topologies $ \mT_{\mu} $ and $ \mT_{\nu} $ on $ \ph $, resp.,
$ S/S^1 $.

\begin{theorem}\label{thm:ss1-esh}
The mapping $ F \! : S/S^1 \to \esh $, $ F([\fii]_S := P_{\fii} $ where
$ \fii \in S $, is a homeomorphism between the topological spaces
$ (S/S^1,\mT_{\nu}) $ and $ (\esh,\mT_0) $.
\end{theorem}
\proof{
The mapping $F$ is bijective. The map
$ h_Q \circ F \circ \nu \! : S \to \R $
where $ h_Q $ is any of the functions given by Eq.~(\ref{p}) and
$ \nu $ is the canonical projection from $S$ onto $ S/S^1 $,
reads explicitly
\[
(h_Q \circ F \circ \nu)(\fii) = h_Q(F([\fii]_S)) = h_Q(P_{\fii})
                              = \tr{P_{\fii}Q}   = \ip{\fii}{Q\fii};
\]
therefore, $ h_Q \circ F \circ \nu $ is continuous. Consequently,
for an open set $ O \subseteq \R $,
\[
(h_Q \circ F \circ \nu)^{-1}(O) = \nu^{-1}(F^{-1}(h_Q^{-1}(O)))
\]
is an open set of $S$. By the definition of the quotient topology
$ \mT_{\nu} $, it follows that $ F^{-1}(h_Q^{-1}(O)) $ is an open set of
$ S/S^1 $. Since the sets $ h_Q^{-1}(O) $, $ Q \in \esh $,
$ O \subseteq \R $ open, generate the weak topology $ \mT_0 $,
$ F^{-1}(U) $ is open for any open set $ U \in \mT_0 $. Hence,
$F$ is continuous.

To show that $F$ is an open mapping, let $ V \in \mT_{\nu} $ be an open
subset of $ S/S^1 $ and let $ [\fii_0]_S \in V $. Since the canonical
projection $ \nu $ is continuous, there exists an $ \varepsilon > 0 $
such that
\begin{equation}\label{nuke}
\nu(K_{\varepsilon}(\fii_0) \cap S) \subseteq V
\end{equation}
where $ K_{\varepsilon}(\fii_0) := \{ \fii \in \hi \, | \,
\no{\fii - \fii_0} < \varepsilon \} $. Without loss of generality
we assume that $ \varepsilon < 1 $.

The topology $ \mT_0 $ is generated by the functions $ h_Q $
according to (\ref{p}); $ \mT_0 $ is also generated by the functions
$ P \mapsto g_Q(P) := \sqrt{h_Q(P)} = \sqrt{\tr{PQ}} $. In consequence,
the set
\[
U_{\varepsilon}
 :=  g_Q^{-1} \left( \, \left] 1 - \tfrac{\varepsilon}{2},
                               1 + \tfrac{\varepsilon}{2} \right[ \, \right)
\cap h_Q^{-1} \left( \, \left] 1 - \tfrac{\varepsilon^2}{4},
                               1 + \tfrac{\varepsilon^2}{4} \right[
                                                                  \, \right)
\]
where $ Q := P_{\fii_0} $ and $ \fii_0 $ and $ \varepsilon $ are specified
in the preceding paragraph, is $ \mT_0 $-open. Using the identity
\[
1 - |\ip{\fii_0}{\fii}|^2 = \no{\fii - \ip{\fii_0}{\fii} \fii_0}^2
\]
where $ \fii \in \hi $ is also a unit vector, we obtain
\begin{eqnarray*}
U_{\varepsilon}
& = & \left\{ P_{\fii} \in \esh \left| \,
        |g_Q(P_{\fii}) - 1| < \tfrac{\varepsilon}{2} \ {\rm and} \
        |h_Q(P_{\fii}) - 1| < \tfrac{\varepsilon^2}{4} \right. \right\}   \\
& = & \left\{ P_{\fii} \in \esh \left| \,
      \bigl| |\ip{\fii_0}{\fii}| - 1 \bigr| < \tfrac{\varepsilon}{2} \
                                                           {\rm and} \
      \bigl| |\ip{\fii_0}{\fii}|^2 - 1 \bigr| < \tfrac{\varepsilon^2}{4}
                                                       \right. \right\}   \\
& = & \left\{ P_{\fii} \in \esh \left| \,
      \bigl| |\ip{\fii_0}{\fii}| - 1 \bigr| < \tfrac{\varepsilon}{2} \
                                                           {\rm and} \,
      \no{\fii - \ip{\fii_0}{\fii} \fii_0} < \tfrac{\varepsilon}{2}
                                                       \right. \right\}.
\end{eqnarray*}
Now let $ P_{\fii} \in U_{\varepsilon} $. Since $ \varepsilon < 1 $, we have
that $ \ip{\fii}{\fii_0} \neq 0 $. Defining the phase factor
$ \lambda := \frac{\ip{\fii}{\fii_0}}{|\ip{\fii}{\fii_0}|} $,
it follows that
\begin{eqnarray*}
\no{\lambda \fii - \fii_0}
& = & \no{\lambda \fii - \lambda \ip{\fii_0}{\fii} \fii_0}
        + \no{\lambda \ip{\fii_0}{\fii} \fii_0 - \fii_0}         \\
& = & \no{\fii - \ip{\fii_0}{\fii} \fii_0}
        + \bigl\| |\ip{\fii_0}{\fii}| \fii_0 - \fii_0 \bigr\|    \\
& < & \tfrac{\varepsilon}{2} + \tfrac{\varepsilon}{2}            \\
& = & \varepsilon.
\end{eqnarray*}
That is, $ P_{\fii} \in U_{\varepsilon} $ implies that
$ \lambda \fii \in K_{\varepsilon}(\fii_0) $; moreover,
$ \lambda \fii \in K_{\varepsilon}(\fii_0) \cap S $.

Taking the result (\ref{nuke}) into account, we conclude that,
for $ P_{\fii} \in U_{\varepsilon} $,
$ [\fii]_S = [\lambda \fii]_S = \nu(\lambda \fii) \in V $. Consequently,
$ P_{\fii} = F([\fii]_S) \in F(V) $. Hence,
$ U_{\varepsilon} \subseteq F(V) $. Since $ U_{\varepsilon} $
is an open neighborhood of $ P_{\fii_0} $, $ P_{\fii_0} $
is an interior point of $ F(V) $. So, for every $ [\fii_0]_S \in V $,
$ F([\fii_0]_S) = P_{\fii_0} $ is an interior point of $ F(V) $, and
$ F(V) $ is a $ \mT_0 $-open set. Hence, the continuous bijective map
$F$ is open and thus a homeomorphism. \qed}

In the following, we identify the sets $ \ph $, $ S/S^1 $, and $ \esh $
and call the identified set the {\it projective Hilbert space
$ \ph $}. However, we preferably think about the elements of
$ \ph $ as the one-dimensional orthogonal projections $ P = P_{\fii} $. On
$ \ph $ then the quotient topologies $ \mT_{\mu} $, $ \mT_{\nu} $, the
weak topologies $ \mT_0 $, $ \mT_w $, $ \mT_1,\ldots,\mT_5 $, $ \mT_s $,
and the metric topologies $ \mT_n $, $ \mT_{\mathrm{tr}} $ coincide. So
we can say that $ \ph $ carries a natural topology $ \mT $;
$ (\ph,\mT) $ is a second-countable Hausdorff space.

For our purposes, it is suitable to represent this topology $ \mT $ as
$ \mT_0 $, $ \mT_n $, or $ \mT_{\mathrm{tr}} $. As already discussed,
the topologies $ \mT_0 $, $ \mT_n $, and $ \mT_{\mathrm{tr}} $ are
canonically related to uniform structures. With respect to the uniform
structure inducing $ \mT_0 $, $ \ph $ is not complete. The uniform
structures related to $ \mT_n $ and $ \mT_{\mathrm{tr}} $ are the same
since they are induced by the equivalent metrics $ \rho_n $ and
$ \rho_{\mathrm{tr}} $; $ (\ph,\rho_n) $ and $ (\ph,\rho_{\mathrm{tr}}) $
are separable complete metric spaces. So $ \mT $ can be defined by
a complete separable metric, i.e., $ (\ph,\mT) $ is a polish space.

\section{The Measurable Structure of $\ph$}\label{sec:meas}

It is almost natural to define a measurable structure on the
projective Hilbert space $ \ph $ by the $ \sigma $-algebra $ \Xi =
\Xi(\mT) $ generated by the $ \mT $-open sets, i.e., $ \Xi $ is the
smallest $ \sigma $-algebra containing the open sets of the natural
topology $ \mT $. In this way $ (\ph,\Xi) $ becomes a measurable
space where the elements $ B \in \Xi $ are the Borel sets of $ \ph
$. However, since the topology $ \mT $ is generated by the
transition-probability functions $ h_Q $ according to Eq.\
(\ref{p}), it is also obvious to define the measurable structure of
$ \ph $ by the $ \sigma $-algebra $ \Sigma $ generated by the
functions $ h_Q $, i.e., $ \Sigma $ is the smallest $ \sigma
$-algebra such that all the functions $ h_Q $ are measurable. A
result due to Misra (1974) \cite[Lemma 3]{mis74} clarifies the
relation between $ \Xi $ and $ \Sigma $. Before stating that result,
we recall the following simple lemma which we shall also use later.

\begin{lemma}\label{lem:sigt-sigb}
Let $ (M,\mT) $ be any second-countable topological space,
$ \mB \subseteq \mT $ a countable base, and $ \Xi = \Xi(\mT) $ the
$ \sigma $-algebra of the Borel sets of $M$. Then
$ \Xi = \Xi(\mT) = \Xi(\mB) $ where $ \Xi(\mB) $ is the
$ \sigma $-algebra generated by $ \mB $; $ \mB $ is
a countable generator of $ \Xi $.
\end{lemma}
\proof{
Clearly, $ \Xi(\mB) \subseteq \Xi(\mT) $. Since every open set
$ U \in \mT $ is the countable union of sets of $ \mB $, it follows that
$ U \in \Xi(\mB) $. Therefore, $ \mT \subseteq \Xi(\mB) $ and consequently
$ \Xi(\mT) = \Xi(\mB) $. \qed}

\begin{theorem}[{\rm Misra}]\label{thm:misra}
The $ \sigma $-algebra $ \Xi = \Xi(\mT) $ of the Borel sets of the
projective Hilbert space $ \ph $ and the $ \sigma $-algbra $ \Sigma $
generated by the transition-probability functions $ h_Q $, $ Q \in \ph $,
are equal.
\end{theorem}
\proof{
Since $ \mT $ is generated by the functions $ h_Q $, the latter are
continuous and consequently $ \Xi $-measurable. Since $ \Sigma $ is
the smallest $ \sigma $-algebra such that the functions $ h_Q $ are
measurable, it follows that $ \Sigma \subseteq \Xi $.

Now, by Lemma \ref{lem:esh}, $ \mT $ is second-countable, and a countable base
$ \mB $ of $ \mT $ is given by the finite intersections of the sets
$ U_{klm} $ according to Eq.\ (\ref{uklm}). Since $ U_{klm} \in \Sigma $,
it follows that $ \mB \subseteq \Sigma $. By Lemma \ref{lem:sigt-sigb}, we conclude that
$ \Xi = \Xi(\mB) \subseteq \Sigma $. Hence, $ \Xi = \Sigma $. \qed}

We remark that our proof of Misra's theorem is much easier than
Misra's proof from 1974. The reason is that we explicitly used the
countable base $ \mB $ of $ \mT $ consisting of $ \Sigma $-measurable sets.

Finally, consider the $ \sigma $-algebra $ \Xi_0 $ in $ \ph $ that is
generated by all $ \mT $-continuous real-valued functions on $ \ph $, i.e.,
$ \Xi_0 $ is the $ \sigma $-algebra of the Baire sets of $ \ph $. Obviously,
$ \Sigma \subseteq \Xi_0 \subseteq \Xi $; so Theorem \ref{thm:misra} implies that
$ \Xi_0 = \Xi $. This result is, according to a general theorem,
also a consequence of the fact that the topology $ \mT $ of
$ \ph $ is metrizable.

Summarizing, our result $ \Sigma = \Xi_0 = \Xi $ manifests that
the projective Hilbert space carries, besides its natural topology
$ \mT $, also a very natural measurable structure $ \Xi $.

\section{The Misra-Bugajski Reduction Map}\label{sec:mb}

The expression $ \tr{WA} $ where $ W \in \sh $ is a density operator
and $A$ a self-adjoint operator, plays a central role
in quantum mechanics. We are going to show how,
for bounded self-adjoint operators $ A \in \bsh $,
this expression can be represented
as an integral over the projective Hilbert space $ \ph $. This
result was first obtained by Misra (1974) \cite{mis74} and
independently by Ghirardi, Rimini and Weber (1976) \cite{ghi76}, and
an elementary construction for the case of a two-dimensional Hilbert
space was discussed by Holevo (1982) \cite{hol82}. The significance
of the representation of quantum expectations on $\ph$ was
elucidated in seminal papers of Bugajski and Beltrametti
\cite{bug91;93a-d,bel95a;b}. Further discussion can be found in
\cite{stu01,bus04}.

\begin{theorem}\label{thm:qc}
For every probability measure $ \mu $ on $ (\ph,\Xi) $,
there exists a uniquely determined density operator $ W_{\mu} \in \sh $
such that, for all $ A \in \bsh $,
\[
\tr{W_{\mu}A} = \int_{\ph} \tr{PA} \ \mu(dP).
\]
\end{theorem}
\proof{
Because of $ |\tr{(P - P_0)A}| \leq \no{P - P_0}_{\mathrm{tr}} \no{A} $
where $ P,P_0 \in \ph $, the function $ P \mapsto \tr{PA} $ on $ \ph $
is continuous w.r.t.\ the metric $ \rho_{\mathrm{tr}} $ and in consequence
$ \mT $-continuous and $ \Xi $-measurable; in addition, because of
$ |\tr{(PA}| \leq \no{A} $, the function is bounded. Hence, the integral
$ \int_{\ph} \tr{PA} \ \mu(dP) $ exists for every probability measure
$ \mu $ on $ \ph $. Moreover, the functional
\[
A \mapsto \phi(A) := \int_{\ph} \tr{PA} \ \mu(dP)
\]
is linear, bounded, and positive. Let $ \{A_n\}_{n \in \N} $
be a sequence of bounded self-adjoint operators satisfying
$ 0 \leq A_n \leq A_{n+1} \leq I $; $ \{A_n\}_{n \in \N} $
converges to some $ A \in \bsh $, $ A \leq I $, with respect to
the weak operator topology, for instance. It follows that,
for all $ P \in \ph $,
\[
0 \leq \tr{PA_n}  \leq \tr{PA_{n+1}} \leq 1
\]
and, writing $ P = P_{\psi} $,
\[
\tr{PA_n} = \langle \psi | A_n\psi \rangle
\rightarrow \langle \psi | A\psi \rangle = \tr{PA}
\]
as $ n \rightarrow \infty $. By the monotone-convergence theorem we
obtain
\[
\phi(A_n) = \int_{\ph} \tr{PA_n} \ \mu(dP)
            \rightarrow \int_{\ph} \tr{PA} \ \mu(dP)
          = \phi(A) ,
\]
i.e., the functional $ \phi $ is normal. Since the normal functionals
on $ \bsh $ can be represented by trace-class operators, there exists an
operator $ W_{\mu} \in \tsh $ such that
\[
\phi(A) = \tr{W_{\mu}A} = \int_{\ph} \tr{PA} \ \mu(dP) .
\]
The operator $ W_{\mu} $ is uniquely determined, self-adjoint, positive,
and, because of $ \tr{W_{\mu}} = \phi(I) = 1 $, of trace $1$, i.e.,
$ W_{\mu} \in \sh $. \qed}

The next theorem summarizes the properties of the mapping
$ \mu \mapsto W_{\mu} $. Remember that the elements of
$ \ph $ are the extreme points of the convex set $ \sh $.

\begin{theorem}\label{thm:mb}
The mapping $ R \! : \spx \to \sh $, $ R(\mu) = W_{\mu} $,
where\linebreak $ \spx $ denotes the convex set of all probability
measures on $ \phx$, has the following properties:
\begin{enumerate}
\item[(a)] $R$ is affine, i.e., for every convex linear combination
           $ \mu = \alpha \mu_1 + (1-\alpha)\mu_2 $ of
           $ \mu_1,\mu_2 \in \spx $, $ 0 \leq \alpha \leq 1 $, we have
           $ W_\mu = \alpha W_{\mu_1} + (1 - \alpha)W_{\mu_2} $;
\item[(b)] $R$ is surjective, but not injective (provided that
           $ \dim \hi \geq 2 $);
\item[(c)] $ R(\mu) = P $, $ P \in \ph $, holds if and only if $ \mu $
           is equal to the Dirac measure $ \delta_P $;
\item[(d)] $R$ maps the Dirac measures on $ \phx $ bijectively onto the
           pure quantum states $ P \in \ph $ and all other probability
           measures on $ \phx $ ``many-to-one'' onto the mixed
           quantum states $ W \in \sh $.
\end{enumerate}
\end{theorem}
\proof{
The first statement is trivial. To prove statement (b), consider any
$ W \in \sh $ and a representation $ W = \sum_{i=1}^{\infty} \alpha_i P_i $
where $ \alpha_i \geq 0 $, $ \sum_{i=0}^{\infty} \alpha_i = 1 $,
$ P_i \in \ph $, and the infinite sum converges
in the trace norm. Define a probability measure
$ \mu \in \spx $ by $ \mu := \sum_{i=1}^{\infty} \alpha_i \delta_{P_i} $
and note that the sum converges in the total-variation norm. Writing
$ \tr{PA} =: f_A(P) $ where $ A \in \bsh $ and $ f_A \in \fpx $,
it follows that
\begin{eqnarray*}
\int_{\ph} \tr{PA} \ \mu(dP)
& = & \dual{\mu}{f_A}
  = \left\langle \sum_{i=1}^{\infty} \alpha_i \delta_{P_i},
                                          f_A \right\rangle              \\
& = & \sum_{i=1}^{\infty} \alpha_i \dual{\delta_{P_i}}{f_A}              \\
& = & \sum_{i=1}^{\infty} \alpha_i \int_{\ph} \tr{PA} \ \delta_{P_i}(dP) \\                        \\
& = & \sum_{i=1}^{\infty} \alpha_i \, \tr{P_iA}                          \\
& = & \tr{WA},
\end{eqnarray*}
which implies $ W = W_{\mu} = R(\mu) $. Hence, $R$ is surjective. Since
every mixed quantum state can be represented in many ways as an infinite
convex linear combination of one-dimensional orthogonal projections,
not necessarily being mutually orthogonal (cf.\ \cite{lud83,bel81}),
let
\[
W = \sum_{i=1}^{\infty} \alpha_i P_i = \sum_{i=1}^{\infty} \beta_i Q_i ,
\qquad \mu_1 := \sum_{i=1}^{\infty} \alpha_i \delta_{P_i} ,
\qquad \mu_2 := \sum_{i=1}^{\infty} \beta_i \delta_{Q_i} ,
\]
where two different representations of any $ W \in \sh \setminus \ph $
have been chosen. Then $ W = R(\mu_1) = R(\mu_2) $ holds, but
$ \mu_1 \neq \mu_2 $; that is, $R$ is not injective.

Since $ R(\delta_P) = P $ is a trivial fact, we have, in order
to prove (c), only to show that $ R(\mu) = P $ implies
$ \mu = \delta_P $. From $ R(\mu) = P $, resp.,
$ \tr{PA} = \int_{\ph} \tr{QA} \ \mu(dQ) $
we obtain, setting $ A = P $,
\[
1 = \int_{\ph} \tr{QP} \ \mu(dQ)
\]
which can be rewritten as
\[
\int_{\ph} (1 - \tr{QP}) \ \mu(dQ) = 0.
\]
Because the integrand is nonnegative, it must vanish almost everywhere. It
follows that
\[
\mu(\{ Q \in \ph \, | \, \tr{QP} = 1 \}) = 1
\]
or, equivalently, $ \mu(\{ P \}) = 1 $. That is, the probability measure
$ \mu $ is concentrated at the point $ P \in \ph $ and consequently
equal to the Dirac measure $ \delta_P $.

Statement (d) is a consequence of (c), (b), and the proof of the fact
that $R$ is not injective. \qed}

Consider now the unique linear extension $ R \! : \mpx \to \tsh $
of the affine mapping $ R \! : \spx \to \sh $. The extended map $R$
is determined by
\begin{equation}\label{mb-map}
\tr{(R\nu)A} = \int_{\ph} \tr{PA} \ \nu(dP)
\end{equation}
where $ \nu \in \mpx $ and $ A \in \bsh $. From
\[
\dual{R\nu}{A} = \int_{\ph} \tr{PA} \ \nu(dP) = \dual{\nu}{f_A}
\]
where $ f_A(P) = \tr{PA} $ it follows that the dual map $R'$ of $R$
w.r.t.\ the considered dualities $ \dual{\mpx}{\fpx} $ and $
\dual{\tsh}{\bsh} $ exists and is given by $ R'A = f_A $. The
existence of $R'$ in this sense means that the range of the usual
adjoint map $ R^* \! : \bsh \to (\mpx)' $ is under $ \fpx $.
According to the discussion in the introduction and the definition
there, $R$ is a reduction map and $ \dual{\spx}{\epx} $ a classical
extension of the quantum statistical model $ \dual{\sh}{\eh} $. We
call the reduction map $R$ given by (\ref{mb-map}) the {\it
Misra-Bugajski map}. The affine mapping $R$ was
introduced by Misra in 1974 \cite{mis74}
who considered it as a new way of defining the notion of quantum
state; it was the late S.~Bugajski who realized that this map
determines a classical extension of the quantum statistical duality
and who initiated a research program to elucidate the physical
significance of this extension---see, e.g.,
\cite{bug91;93a-d,bel95a;b}.

The adjoint $R'$ of the Misra-Bugajski map $R$ associates the
quantum mechanical effects $ A \in \eh $ with the classical effects
$ R'A = f_A \in \epx $. However, except for the trivial cases $ A =
0 $ or $ A = I $, such a function $ f_A $, $f_A(P)= \tr{PA} $, is
never the characteristic function $ \chi_B $ of some set $ B \in \Xi
$; that is, the functions $ f_A $ describe unsharp (fuzzy) effects.

\section{The Representation of Classical Extensions of Quantum Mechanics}\label{sec:cextq}

Now we are going to show that every classical extension of quantum
mechanics is essentially given by the Misra-Bugajski reduction map.
This result was conjectured in \cite{bus04}, and the proof given
here takes up elements of a very rough sketch given there.

Assume a classical extension on a measurable space $ \os $ is given by the
linear maps $ R \! : \mos \to \tsh $ and $ R' \! : \bsh \to \fos $. Then,
for $\mu\in\sos$ and $A\in\bsh$, we have
\begin{equation}
\tr{(R\mu)A} = \dual{R\mu}{A} = \dual{\mu}{R'A}
             = \int_{\Omega} R'A \, d\mu; \label{R-mu}
\end{equation}
setting $ \mu = \delo $ where $ \delo $ denotes the Dirac measure of a point
$ \omega \in \Omega $, we obtain
\begin{equation}
(R'A)(\omega) = \tr{(R\delo)A}. \label{R-A}
\end{equation}
Hence,
\begin{equation}
\tr{(R\mu)A} = \int_{\Omega} \tr{(R\delo)A \ \mu(d\omega)}. \label{tr-int}
\end{equation}
To prove our main result, Theorem \ref{thm10} below, we need several lemmata.

\begin{lemma}\label{lem1}
For $ P \in \ph $, the set  $ \{ \omega \in \Omega \, | \, R\delo = P \} $
is measurable. If $P=R\mu$, then
\[
\mu(\set{\omega\in\Omega}{R\delo = P}) = 1.
\]
In particular, for every $P\in\ph$ there exists an $\omega\in\Omega$
such that $R\delo=P$.
\end{lemma}
\proof{
Let $ E_P := \{ \omega \in \Omega \, | \, R\delo = P \} $. Since
the statement $ R\delo = P $ is equivalent to $ \tr{(R\delo)P} = 1 $,
it follows that
\[
E_P = \{ \omega \in \Omega \, | \, \tr{(R\delo)P} = 1 \}.
\]
Setting $ A = P $ in Eq.\ (\ref{R-A}), we see that the function
$ P \mapsto \tr{(R\delo)P} $ is measurable; therefore, the set $ E_P $
is measurable. Setting $ P = R\mu $ and $ A = P $ in Eq.~(\ref{tr-int}),
we obtain
\[
\int_{\Omega} \tr{(R\delo)P} \ \mu(d\omega) = 1
\]
which can be rewritten as
\[
\int_{\Omega} (1-\tr{(R\delo)P}) \ \mu(d\omega)=0.
\]
Since the integrand is nonnegative, it must vanish almost everywhere. Hence,
\[
\mu(E_P) = \mu(\set{\omega\in\Omega}{1 - \tr{(R\delo)P} = 0}) = 1.
\]
Because $R$ is surjective, every $ P \in \ph $ is of the form
$ P = R\mu $. Then $ \mu(E_P) = 1 $ implies that $ E_P $ is not empty.} \qed

\begin{lemma}\label{lem3}
Let $P_n\in\ph$, $n\in\mathbb N$, and assume that, for some $W_0\in\sh$,
\begin{equation}
\lim_{n\to\infty}\tr{W_0P_n}= 1. \label{W0-lim}
\end{equation}
Then there exists an element $P\in\ph$ such that
$\lim_{n\to\infty}\no{P_n-P}=0$; moreover, $W_0=P$.
\end{lemma}
\proof{For each $n\in\mathbb N$, let $\fii_n$ be a unit vector in the range
of $P_n$, and write $P_n=P_{\fii_n}$. Since $\no{\fii_n}= 1$, the weak
compactness of the unit sphere of $\hi$  entails that there is a subsequence
$\{\fii_{n_j}\}_{j \in \N}$ of $\{\fii_n\}_{n\in\mathbb N}$ converging
weakly to some $ \psi \in \hi $, $\no{\psi}\le 1$.

Let $W$ be any element of $\sh$. We show that
$ \tr{WP_{\fii_{n_j}}} \to \tr{(W\kb{\psi}{\psi})} $ as
$ j \to \infty $. The density operator can be written as
$ W = \sum_{i=1}^{\infty} \alpha_i P_{\chi_i} $ where
$ \alpha_i \geq 0 $, $\sum_{i=1}^\infty\alpha_i=1$, $ \chi_i \in \hi $, and
$ \no{\chi_i} = 1 $. Choose $ \varepsilon>0 $ and a number $ N_0 \in \N $
such that $ \sum_{i=N_0+1}^\infty \alpha_i < \frac\varepsilon 4 $. Since
the sequence $ \{ \fii_{n_j} \}_{j \in \N} $ converges weakly to
$ \psi $, there is an integer $ J(\varepsilon) $ such that for all
$ j \geq J(\varepsilon) $ and all $ i = 1,\ldots,N_0 $,
\[
|\ip{\chi_i}{\fii_{n_j}}|^2-|\ip{\chi_i}{\psi}|^2 < \tfrac\varepsilon 2.
\]
It follows that, for all $ j \geq J(\varepsilon) $,
\begin{eqnarray*}
\left| \tr{WP_{\fii_{n_j}}} - \tr{(W\kb{\psi}{\psi})} \right|
&  =   &    \left| \sum_{i=1}^\infty \alpha_i |\ip{\chi_i}{\fii_{n_j}}|^2
         -  \sum_{i=1}^\infty \alpha_i |\ip{\chi_i}{\psi}|^2 \right|      \\
& \leq &    \left|  \sum_{i=1}^{N_0} \alpha_i
            \left( |\ip{\chi_i}{\fii_{n_j}}|^2 - |\ip{\chi_i}{\psi}|^2
                                                     \right) \right|
         + 2\sum_{i=N_0+1}^\infty\alpha_i                                 \\
&  <  &    \tfrac\varepsilon 2 + \tfrac\varepsilon 2                      \\
&  =  &    \varepsilon.
\end{eqnarray*}
Hence,
\begin{equation}
\lim_{j\to\infty}\tr{WP_{\fii_{n_j}}}=\tr{(W\kb{\psi}{\psi})}. \label{W-lim}
\end{equation}
For $ W = W_0 $, Eqs.\ (\ref{W0-lim}) and (\ref{W-lim}) imply that
\[
\tr{(W_0\kb{\psi}{\psi})} = 1.
\]
So $\psi\ne 0$; defining $ \Psi := \frac{\psi}{\no{\psi}} $, we obtain
$ \no{\psi}^2 \tr{W_0P_\Psi} = 1 $. It follows immediately that
$ \no{\psi}=1 $ and $ \tr{W_0P_\Psi} = 1 $. Hence,
$ \tr{W_0P_{\psi}} = \ip{\psi}{W_0\psi} = 1 $, that is,
$W_0$ has the eigenvalue $1$ with multiples of $ \psi $ as eigenvectors,
i.e., $ W_0 = P_\psi =: P $.

It remains to show that $\no{P_n-P}\to 0$ as $n\to\infty$. From
(\ref{W0-lim}) and $W_0=P$ it follows that $ \tr{PP_n} \to 1 $ as $
n \to \infty $. But this is, according to Theorem
\ref{thm:esh-dist}, part~(a), equivalent to
\[
\no{P_n - P}^2 = 1 - \tr{PP_n} \to 0
\]
as $ n \to \infty $.} \qed

It can be shown that the norm convergence of a sequence $ \{ P_n
\}_{n \in \N} $ in $ \ph $, $ P_n = P_{\fii_n}, $ to $ P = P_{\psi}
\in \ph $ entails the existence of a subsequence
$\{\fii_{n_j}\}_{j\in\mathbb N}$ of $\{\fii_n\}_{n\in\mathbb N}$
such that $\lim_{j\to\infty}\no{\fii_{n_j}-e^{i\alpha}\psi}=0$ with
some $\alpha\in\mathbb R$. The example
\[
\fii_n: = e^{in\pi}\psi=(-1)^n\psi, \quad
\no{P_{\fii_n} - P_\psi} \to 0 \ {\rm as} \ n \to \infty
\]
shows that convergence at the level of vectors can follow only for
a subsequence. Concerning the sequences $ \{ \fii_n \}_{n \in \N} $ and
$ \{ \fii_{n_j} \}_{j \in \N} $ introduced at the beginning
of the preceding proof, it finally turns out that the subsequence
$ \{ \fii_{n_j} \}_{j \in \N} $ is even norm-convergent (which
is not essential for the proof), however, the restriction of
$ \{ \fii_n \}_{n \in \N} $ to a subsequence is essential.

\begin{lemma}\label{lem5}
Let
\[
\omti := \set{\omega \in \Omega}{R\delo \in \ph}
       = \set{\omega \in \Omega}{\tr{(R\delo)P} = 1 \ {\rm for \ some}
                                                       \ P \in \ph}.
\]
Then $\omti$ is a measurable subset of $\Omega$.
\end{lemma}
\proof{Let $ \{ P_m \}_{m\in\N} $ be a $\no{\cdot}$-dense sequence in $\ph$
and let
\[
\Omega_{mn}: = \left\{ \omega \in \Omega \left| \,
                       \tr{(R\delo)P_m} > 1 - \tfrac{1}{n} \right. \right\}
\]
where $n\in\N$. We show that
\begin{equation}\label{omti-capcup}
\omti=\bigcap_{n\in\N}\bigcup_{m\in\N}\Omega_{mn}
\end{equation}
holds.

Let $\omega\in\omti$ and $R\delo=P$, i.e., $\tr{(R\delo)P}=1$. For every
$n\in\N$ there exists a member $P_m$ of the dense sequence satisfying
$\no{P_m-P}<\frac 1n$, in consequence,
\[
1-\tr{(R\delo)P_m} = \left|\tr{(R\delo)P_m}-\tr{(R\delo)P}\right|
\le \no{R\delo}_{\mathrm{tr}} \no{P_m-P} <\tfrac 1n;
\]
that is, $\tr{(R\delo)P_m}>1-\frac 1n$. Hence,
$\omega\in\bigcap_{n\in\N}\bigcup_{m\in\N}\Omega_{mn}$.

Conversely, assume
$\omega\in\bigcap_{n\in\N}\bigcup_{m\in\N}\Omega_{mn}$. Then
for every $n\in\N$ there is an $m\in\N$ with $\omega\in\Omega_{mn}$. In
other words, for every $n\in\N$ there exists at least one $P_m$ such that
$\tr{(R\delo)P_m}>1-\frac 1n$. Let $P_{m_n}$ be such a $P_m$. Then it holds
true that $ 1 - \frac{1}{n} < \tr{(R\delo)P_{m_n}} \le 1 $, which implies
that
\[
\tr{(R\delo)P_{m_n}} \to 1
\]
as $ n \to \infty $. By virtue of Lemma \ref{lem3}, this entails
$R\delo=P\in\ph$, that is, $\omega\in\omti$. Thus, Eq.~(\ref{omti-capcup})
has been proved.

Due to the measurability of the functions
$ \omega \mapsto (R'A)(\omega) = \tr{(R\delo)A} $ for $A\in\bsh$,
the sets $\Omega_{mn}$ are measurable; from  Eq.~(\ref{omti-capcup})
one then concludes that $\omti\in\Sigma$.} \qed

Next we shall redefine our reduction map $ R \! : \mos \to \fos $
w.r.t.\ the measurable space $ (\omti,\sigti) $ where
$ \sigti := \Sigma \cap \omti $ (since $ \omti $ is measurable,
we have that $ \sigti = \{ E \in \Sigma \, | \, E \subseteq \omti \}
\subseteq \Sigma $). To that end, we introduce
\[
\mN := \bigl\{ \nu \in \mos \, \bigl| \, \nu(E)=0, \ E \in \Sigma,
                            \ E \subseteq \Omega \setminus \omti \bigl\}
\]
and
\begin{eqnarray*}
\sn & := & {\bigl\{ \mu \in \sos \, \bigl| \, \mu(\Omega \setminus \omti)
                                                         = 0 \bigl\}}
       =    \bigl\{ \mu \in \sos \, \bigl| \, \mu(\omti) = 1 \bigl\}      \\
    &  = &  \mN \cap \sos.
\end{eqnarray*}
The set $\mN$ is a norm-closed subspace of $ \mos $, and $ \sn $ is
a norm-closed face of $ \sos $. Moreover, $ (\mN,\sn) $ is a base-normed
Banach space with closed positive cone; we do not need these results
here. The spaces $ \mN $ and $ \mosti $ are canonically related by
the linear map $ J \! : \mN \to \mosti $ defined by
\[
\nu \mapsto \nuti = J\nu := \nu\vert_{\sigti}
\]
where $ \nu\vert_{\sigti} $ denotes the restriction of
$ \nu $ to $ \sigti $; $ J $ is a linear isomorphism preserving
norm and order. The inverse $ J^{-1} $ is given by
\[
\nuti \mapsto \nu = J^{-1}\nuti, \quad \nu(A) = \nuti(A \cap \omti)
\]
where $ A \in \Sigma $. We shall only use that $J$ is a linear
isomorphism.---In the context of the following theorem, $ \deloti $
denotes the restriction of the Dirac measure $ \delo $, defined on $
\Sigma $ and concentrated at $ \omega \in \omti $, to $ \sigti $.

\begin{theorem}\label{thm6}
Let a linear map $ \rti \! : \mosti \to \tsh $ be defined according to
$ \rti\nuti := R\nu $ where $ J\nu = \nuti $, i.e., $ \rti = RJ^{-1} $. Then
\begin{enumerate}
\item[(i)] $ \rti\sosti = \sh $;
\item[(ii)] $ \rti $ is
            $ \sigtop{\mosti,\fosti} $-$ \sigtop{\tsh,\bsh}
            $-continuous;
\item[(iii)] $ \bigl\{ \rti\deloti \bigl| \, \omega \in \omti \bigr\}
             = \ph $.
\end{enumerate}
That is, $ \rti $ is a reduction map with the additional property (iii).
\end{theorem}
\proof{
We prove statement (iii) first. By the definition of
$ \omti $ in Lemma \ref{lem5} it is clear that
$ \bigl\{ R\delo \bigl| \, \omega \in \omti \bigl\} \subseteq \ph $. Let
$ P \in \ph $, then by virtue of Lemma \ref{lem1} there exists an
$ \omega \in \Omega$ such that $ R\delo = P $; again by the definition of
$ \omti $, $ \omega \in \omti $. Hence,
$ \bigl\{ R\delo \bigl| \, \omega \in \omti \bigl\} = \ph $; furthermore,
$ R\delo = \rti\deloti $ for $ \omega \in \omti $.

We have $ \rti\sosti = R\sn \subseteq R\sos = \sh $, thus
$ \rti\sosti \subseteq \sh $. Let $ W \in \sh $, and write
$ W = \sum_{i=1}^{\infty} \alpha_i P_i $ where $ \alpha_i \geq 0 $,
$ \sum_{i=1}^{\infty} \alpha_i = 1 $, and $ P_i \in \ph $. Defining
$ \muti: = \sum_{i=1}^\infty \alpha_i \tilde{\delta}_{\omega_i} $ where
$ P_i = \rti \tilde{\delta}_{\omega_i} $ and $ \omega_i \in \omti $,
we obtain a probability measure $ \muti \in \sosti $. It follows that
\[
\rti\muti = \sum_{i=1}^{\infty} \alpha_i \rti \tilde{\delta}_{\omega_i}
          = \sum_{i=1}^{\infty} \alpha_i P_i = W;
\]
for this conclusion we have used that the sums converge in the
respective norms and $ \rti $ is norm-continuous, the latter due to
the linearity of $ \rti $ and the property $ \rti\sosti \subseteq
\sh $ already shown above. Hence, $ \rti\sosti = \sh $.

Taking account of $ \nu = J^{-1}\nuti \in \mN $ for $ \nuti \in \mosti $
and using the abbreviation $ f_A := R'A $ where $ A \in \bsh $,
we obtain that
\begin{eqnarray*}
\dual{\rti\nuti}{A}
& = & \tr{(\rti\nuti)A} = \tr{(R\nu)A}                                   \\
& = & \int_{\Omega}R'A \, d\nu = \int_\Omega f_A\chi_{\omti} \, d\nu     \\
& = & \int_{\omti} f_A \, d\nu = \int_{\omti} f_A\vert_{\omti} \, d\nuti \\
& = & \dual{\nuti}{f_A\vert_{\omti}}\\
& = & \dual{\nuti}{\rti'A};
\end{eqnarray*}
that is, the map $ \rti' \! : \bsh \to \fosti $ being dual to $ \rti $
w.r.t.\ the dualities $ \dual{\tsh}{\bsh} $ and $ \dual{\mosti}{\fosti} $
exists.} \qed

In the sequel we omit the tilde notation and understand by
$R:\mos\to\tsh$ a linear map with the properties (i)-(iii)
of Theorem \ref{thm6}. We have again that
\begin{equation}
\tr{(R\mu)A} = \int_{\Omega} R'A \, d\mu
             = \int_{\Omega} \tr{(R\delo)A \ \mu(d\omega)} \label{tr-int2}
\end{equation}
holds for all $\mu\in\sos$ and $A\in\bsh$
(cf.\ Eqs.\ (\ref{R-mu})-(\ref{tr-int})). Moreover, now the equality
\begin{equation}
\ph = \{ R\delo | \, \omega \in \Omega \} \label{ph-R}
\end{equation}
is satisfied.

\begin{lemma}\label{lem7}
Let $\mT$ be the natural topology of $\ph$ and $\Xi=\Xi(\mT)$ the
$\sigma$-algebra of the Borel sets of $\ph$. The mapping
$ i \! : \Omega \to \ph $ defined by $ i(\omega) := R\delo $
is $\Sigma$-$\Xi$-measurable.
\end{lemma}
\proof{ The topology $ \mT $ is generated by the functions $ h_Q $
defined by Eq.~(\ref{p}). According to
\[
h_Q(i(\omega)) = \tr{i(\omega)Q} = \tr{(R\delo)Q} = (R'Q)(\omega)
\]
where Eq.\ (\ref{R-A}) has been taken into account, the functions $
h_Q \circ i $ are $ \Sigma $-mea\-surable.

Let $ O \subseteq \R $ be an open set. Then
\begin{equation}
U: = h_Q^{-1}(O) \in \mT. \label{U-T}
\end{equation}
From the measurability of the functions $ h_Q \circ i $ it follows that
\[
i^{-1}(U) = i^{-1}(h_Q^{-1}(O)) = (h_Q \circ i)^{-1}(O) \in \Sigma;
\]
that is, for all $U$ of the form (\ref{U-T}) we have
\begin{equation}
i^{-1}(U)\in\Sigma. \label{i-U-Sig}
\end{equation}
According to Lemma \ref{lem:esh}, for a sequence $ \{ Q_k \}_{k \in \N} $ being dense in
$ \ph $, a sequence $ \{ q_l \}_{l \in \N} $ of numbers being dense in
$ [0,1] $, and $ m \in \N $, the finite intersections of the sets
\[
U_{klm}
 = h_{Q_k}^{-1} \left( \, \left] q_l - \tfrac{1}{m},
                                 q_l + \tfrac{1}{m} \right[ \, \right)
\]
form a countable basis $ \mB $ of the topology $ \mT $ of $ \ph $. From this
and from (\ref{i-U-Sig}) we obtain that
\[
i^{-1}(U) \in \Sigma
\]
for all $ U \in \mB $.

In virtue of Lemma \ref{lem:sigt-sigb}, the countable basis $\mB$ of $\mT$ is a (countable)
generator of $\Xi(\mT)$. Since $i^{-1}(U)\in\Sigma$ for all sets
$U$ of a generator of $\Xi=\Xi(\mT)$, the mapping $i$ is
$\Sigma$-$\Xi$-measurable.} \qed

By virtue of Eq.\ (\ref{ph-R}), $i$ is a surjective measurable mapping.

\begin{theorem} \label{thm10}
Any reduction map $R$ with the property 
$\{ R\delo | \, \omega \in \Omega \}=\ph$ can be represented
according to
\begin{equation}
\tr{(R\mu)A} = \int_{\Omega} \tr{PA} \ (\mu \circ i^{-1})(dP) \label{R-mui}
\end{equation}
where $\mu\in\sos$, $A\in\bsh$, $ i \! : \Omega \to \ph $ is the mapping
$\omega\mapsto i(\omega)=R\delo$, and $ \mu \circ i^{-1} $
the image measure.
\end{theorem}
\proof{
The claim follows from (\ref{tr-int2}), Lemma \ref{lem7}, and
the transformation theorem for integrals:
\begin{eqnarray*}
\tr{(R\mu)A} & = & \int_{\Omega} \tr{(R\delo)A} \ \mu(d\omega)
               =   \int_{\Omega} \tr{i(\omega)A} \ \mu(d\omega)      \\
             & = & \int_{\Omega} \tr{PA} \ (\mu \circ i^{-1})(dP).
                                                \ \ \square
\end{eqnarray*}
} 

Given any reduction map $ R \! : \mos \to \tsh $, every density
operator $ W \in \sh $ is the image of some probability measure $
\mu \in \sos $, i.e., $ W = R\mu $. Theorem \ref{thm10} now states
that, after removing the redundant $ \omega \in \Omega $ for which $
R \delo \not\in \ph $, $W$ is the {\it weak integral}
\begin{equation}
R\mu = \int_{\ph} P \ (\mu \circ i^{-1})(dP) \label{R-muiw}
\end{equation}
of the elements $ P \in \ph $ (i.e., of the identity map of $ \ph $)
w.r.t.\ the probability measure $ \mu \in \spx $. The classical {\it
sample space $ \os $} can be replaced by the {\it phase space $ \phx
$} (for the interpretation of $ \ph $ as a phase space, see Section
\ref{sec:int}), Eqs.\ (\ref{R-mui}) and (\ref{R-muiw}) show the
central role of $ \ph $. Comparing Eq.\ (\ref{R-mui}) with Eq.\
(\ref{mb-map}), the latter specifying the Misra-Bugajski map $
R_{MB} $, we obtain
\begin{equation}\label{r-rmb}
R\mu = R_{MB}(\mu \circ i^{-1}).
\end{equation}
If the surjective measurable map $i$ also transforms the measurable sets of
$ \Sigma $ into measurable sets of $ \Xi $, then every probability measure
$ \mu' \in \spx $ is of the form $ \mu' = \mu \circ i^{-1} $. In this case
$R$ can be replaced by $ R_{MB} $; in the case where not every $ \mu' $ is
of the form $ \mu \circ i^{-1} $, $R$ can be restated as some restriction of
$ R_{MB} $. Summarizing, every classical extension of quantum mechanics is
essentially given by the Misra-Bugajski reduction map; therefore,
$ R_{MB} $ is distinguished under all reduction maps.

However, the examples presented in the next section show that the
mapping $i$ is necessary for the statement of Theorem \ref{thm10}
even if $ \Omega = \ph $.

\section{Examples}\label{sec:ex}

The following examples of reduction maps are also of interest by themselves.

\begin{ex}\label{ex1} {\rm
Let $ \mK $ be an infinite-dimensional closed subspace of the Hilbert space
$ \hi $, $ V \! : \hi \to \hi $ a partial isometry satisfying
$ V\mK = \hi $ and $ V{\mK}^{\perp} = \{ 0 \} $, and let
$ \pk := \{ P \in \ph \, | \, P = P_{\fii}, \|\fii\| = 1, \fii \in \mK \} $
($ \pk $ can be identified with the projective Hilbert space associated with
the Hilbert space $ \mK $). Using the general information given in the
paragraph after the proof of Lemma \ref{lem3}, one easily proves that $ \pk $ is
a norm-closed subset of $ \ph $; therefore, $ \pk $ is $ \Xi $-measurable,
and the following integral in (\ref{W-mu1}) makes sense. In fact,
according to
\begin{equation}
\tr{W_\mu A} = \int_{\pk} \tr{VPV^* \! A} \ \mu(dP) \label{W-mu1}
\end{equation}
where $ A \in \bsh $, for each probability measure $ \mu \in \spx $
concentrated on $ \pk $, i.e., $ \mu(\pk) = 1 $, a density operator
$ W_{\mu} \in \sh $ is defined. We can identify the set of these
probability measures with $ \mS(\pk,\Xi_{\mK}) $ where
$ \Xi_{\mK} := \Xi \cap \pk =\{ B \in \Xi \, | \, B \subseteq \pk \}
\subseteq \Xi $. Moreover, the affine mapping $ \mu \mapsto W_\mu $ can
be extended to a reduction map $ R \! : \mM_{\R}(\pk,\Xi_{\mK}) \to \tsh $;
$R$ maps the Dirac measures of $ \mS(\pk,\Xi_{\mK}) $ bijectively onto
$ \ph $, namely, $ R\delp = VPV^* $, $ P \in \pk $.

Setting $ (\Omega,\Sigma) := (\pk,\Xi_{\mK}) $, it follows from Lemma \ref{lem1}
that, for $ Q \in \ph $ and any $ \mu \in \mS(\pk,\Xi_{\mK}) $,
$ R\mu = Q $ if and only if $ \mu = \delp $ with $ P = V^*QV $. Furthermore,
we have for the set $ \omti $ introduced in Lemma \ref{lem5} and for the mapping
$ i \! : \omti \to \ph $ of Lemma \ref{lem7} that $ \omti = \Omega $ and
$ i(P) = R\delp = VPV^* $. In particular, if $ \mK = \hi $ (where
$ \hi $ need not be infinite-dimensional) and $V$ is a unitary operator,
then $ \Omega = \ph = \omti $ and $ i(P) = VPV^* $.
}
\end{ex}

\begin{ex}\label{ex2}{\rm
Letting $ \mK $, $V$, and $ \pk $ as in the preceding example, then
for each probability measure $ \mu \in \spx $ a density operator
$ W_\mu \in \sh $ is defined according to
\begin{equation}
\tr{W_\mu A} =   \int_{\pk} \tr{VPV^* \! A} \ \mu(dP)
               + \int_{\ph\setminus\pk} \tr{PA} \ \mu(dP) \label{W-mu2}
\end{equation}
where $ A \in \bsh $ and $ \ph \setminus \pk $ is the set-theoretical
complement of $ \pk $. Note that $ \mu $ is a probability measure on
$ \ph $ whereas in the preceding example $ \mu $ is a probability measure on
$ \pk $. The affine mapping $ \mu \mapsto W_{\mu} $ given by (\ref{W-mu2})
can be extended to a reduction map $ R \! : \mpx \to \tsh $; $R$ maps
the Dirac measures of $ \spx $ onto $ \ph $, partially two-to-one:
\[
R\delp = \left\{ \begin{array}{ccl}
                  VPV^* & {\rm if} & P \in \pk                \vspace{1mm}\\
                   P    & {\rm if} & P \in \ph \setminus \pk.
                 \end{array} \right.
\]
In fact, from $ R\delp = Q $ it follows that $ P = V^*QV $ if $ Q \in \pk $,
and $ P = V^*QV $ or $ P = Q $ if $ Q \in \ph \setminus \pk $. By Lemma \ref{lem1},
$ R\mu = Q $ for any $ \mu \in \spx $ is equivalent to
$ \mu = \delta_{V^*QV} $ if $ Q \in \pk $, resp., to
$ \mu = \alpha \delta_{V^*QV} + (1 - \alpha) \delta_Q $,
$ 0 \leq \alpha \leq 1 $, if $ Q \in \ph \setminus \pk $.

Setting $ \os := \phx $, we obtain
$ \omti = \Omega $ and $ i \! : \omti \to \ph $, $ i(P) = R\delp
= \chi_{\pk}(P) \, VPV^* + \chi_{\ph \setminus \pk}(P) \, P $ where
$ \chi_{\pk} $, for instance, is the characteristic function of the set
$ \pk $.
}
\end{ex}

\begin{ex}\label{ex3} {\rm
Now let $ \mK $ be an infinite-dimensional closed subspace of $ \hi $ with
an infinite dimensional orthocomplement $ \mK^\perp $ and let $ V_1 $ and
$ V_2 $ be partial isometries satisfying
\[
\begin{array}{lccclcl}
V_1\mK       & = & \hi, & \quad & V_1\mK^\perp & = & \{0\}   \vspace{1mm}\\
V_2\mK^\perp & = & \hi, & \quad & V_2\mK\      & = & \{0\}.
\end{array}
\]
Then each probability measure $ \mu \in \spx $ determines a density operator
$ W_\mu \in \sh $ according to
\begin{equation}
\tr{W_\mu A} = \int_{\ph} \tr{(V_1PV_1^* + V_2PV_2^*)A} \ \mu(dP)
                                                          \label{W-mu3}
\end{equation}
where $ A \in \bsh $. The affine mapping $ \mu \mapsto W_{\mu} $ given by
(\ref{W-mu3}) again extends to a reduction map $ R \! : \mpx \to \tsh $;
$R$ maps the Dirac measures of $ \spx $ onto the quantum states
\begin{eqnarray*}
R\delp & = &    V_1PV_1^* + V_2PV_2^*
         =     \kb{V_1\fii}{V_1\fii} + \kb{V_2\fii}{V_2\fii}   \\
       & = &   \no{\chi_1}^2 P_{\frac{\chi_1}{\no{\chi_1}}}
             + \no{\chi_2}^2 P_{\frac{\chi_2}{\no{\chi_2}}}
\end{eqnarray*}
where $ P = P_{\fii} $, $ \chi_1 := V_1\fii $, $ \chi_2 := V_2\fii $, and
$ \no{\chi_1}^2 + \no{\chi_2}^2 = 1 $. In general, the states
$ R\delp $ are mixed; $ R\delp \in \ph $ is equivalent to
$ P = P_{\fii} $ with $ \fii = a\fii_1 + b\fii_2 $,
$ \fii_1 \in \mK $, $ \fii_2 \in \mK^{\perp} $,
$ \no{\fii_1} = \no{\fii_2} = 1 $, $ a,b \in \C $, $ |a|^2 + |b|^2 = 1 $,
and $ V_1\fii_1 = V_2\fii_2 $. In particular, for each $ Q \in \ph $,
there is one unit vector $ \fii_1 \in \mK $ and one unit vector
$ \fii_2 \in \mK^{\perp} $ such that
$ R\delta_{P_{\fii_1}} = R\delta_{P_{\fii_2}} = Q $,
$ \fii_1 $ and $ \fii_2 $ are uniquely determined up to phase factors. Let
$ \mK_Q $ be the two-dimensional subspace of $ \hi $ that is spanned by
$ \fii_1 $ and $ \fii_2 $ and let $ \mP(\mK_Q) := \{ P \in \ph \, | \,
P = P_{\fii}, \no{\fii} = 1, \fii \in \mK_Q \} $. Then $ R\delp = Q $
if and only if $ P \in \mP(\mK_Q) $, and by Lemma \ref{lem1}, $ R\mu = Q $ for any
$ \mu \in \spx $ if and only if $ \mu $ is concentrated on $ \mP(\mK_Q) $,
i.e., $ \mu(\mP(\mK_Q)) = 1 $.

It follows that $ \mK_{Q_1} \cap \mK_{Q_2} = \{ 0 \} $ as well as
$ \mP(\mK_{Q_1}) \cap \mP(\mK_{Q_2}) = \emptyset $ for $ Q_1 \neq Q_2 $
and that $ \bigcup_{Q \in \ph} \mK_Q \neq \hi $ as well as
$ \bigcup_{Q \in \ph} \mP(\mK_Q) \neq \ph $. Writing $ \os := \phx $,
we obtain $ \omti = \{ P \in \ph \, | \, P \in \mP(\mK_Q) \ {\rm for} \
{\rm some} \ Q \in \ph \} = \bigcup_{Q \in \ph} \mP(\mK_Q) $,
$ \omti \neq \Omega $, and $ i \! : \omti \to \ph $,
$ i(P) = R\delp = V_1PV_1^* + V_2PV_2^* $.
}
\end{ex}

\section{Physical Interpretation}\label{sec:int}


Interpreting the bounded self-adjoint operators on $\hi$ as quantum
observables with real values, the expectation value of $ A \in \bsh
$ in the state $ W \in \sh $ is given by $ {\rm tr} \, WA $.
Analogously, if $ \Omega $ is a classical phase space with the Borel
structure $ \Sigma $, the states are described by the probability
measures on $ \Omega $ and the observables by the (bounded)
measurable functions on $ \Omega $; the expectation value of a
classical observable $ f \in \fos $ in the state $ \mu \in \sos $ is
$ \int f d\mu $. According to Theorems \ref{thm:qc} and
\ref{thm:mb}, each $ W \in \sh $ is of the form $ W = R\mu = W
_{\mu} $, $ \mu $ being some probablity measure on $ \Omega = \ph $.
That is, for every $ W \in \sh $ there exists a probability measure
$ \mu \in \spx $ such that for all $ A \in \bsh $, $ A = A^* $,
\begin{equation}\label{qc}
{\rm tr} \, WA = \int_{\ph} f_A d\mu
\end{equation}
holds where $ f_A $ is the function $ P \mapsto f_A(P) = {\rm tr} \, PA $
on $\ph$. Viewing the projective Hilbert space as a classical phase space,
this result means that the quantum states can be seen as classical states
and the quantum observables as classical ones where the expectation values
can be expressed in classical terms. However, the injective map
$ A \mapsto f_A $ is not surjective, as is easily seen. That is, not all
classical observables on $\ph$ represent quantum ones, which is related to
the fact that the quantum states $W$ correspond to the equivalence classes
$ R^{-1}(\{W\}) $ of classical states, each member of an equivalence class
giving the same quantum mechanical expectation values.

Taking up the notion of quantum statistical model reviewed 
in the introduction, the result (\ref{qc}) can,
much more fundamentally, be interpreted in terms of probabilities if
the operators $A$ are specified to be effects; in that case,
$ {\rm tr} \, WA $ is interpreted to be the probability for the occurrence
of `yes' of the effect $A$ in the state $W$. Eq.~(\ref{qc}) then states
that the quantum mechanical effects $ A \in \eh $ can classically
be described by measurable functions taking values between the numbers
$0$ and $1$, i.e., by the classical effects $ f_A \in \epx $. In the context
of classical probability theory, such effects can be interpreted
as ``unsharp'' measurements of events, these being
the classical analogs of the quantum mechanical effects
and extending probability theory to {\it operational} or {\it fuzzy
probability theory} (cf.\ \cite{gud79,stu86,bug96,gud98}). Again,
the map $ A \mapsto f_A $, $ 0 \leq A \leq 1 $, into the measurable
functions $f$ on $\ph$, $ 0 \leq f \leq 1 $, is injective, but not
surjective. In particular, the orthogonal projections, describing
the ideal quantum mechanical yes--no measurements, are not mapped
onto the characteristic functions, except for the trivial cases;
the ``sharp'' classical events do not correspond to any quantum
mechanical effects.

In general, quantum observables with values in some space $M$,
$ (M,\Upsilon) $ being a measurable space, are operationally described by
{\it positive operator-valued measures (POVMs)} $ F \! : \Upsilon \rightarrow \bsh $, $ b \mapsto F(b) $,
$ 0 \leq F(b) \leq 1 $;
\[
b \mapsto {\rm tr} \, WF(b)
\]
is the probability distribution of the observable $F$ in the state
$ W \in \sh $. The analogous classical concept is that of {\it fuzzy
random variables} which generalizes the usual concept of random variables
(cf.\ \cite{stu86,sin92,bug98,gud98}). Given a classical sample or
phase space $ (\Omega,\Sigma) $ and a space $ (M,\Upsilon) $ of
possible measurement results, a fuzzy random variable is a Markov kernel
$ K \! : \Omega \times \Upsilon \rightarrow [0,1] $, i.e., for each
$ b \in \Upsilon $, $ K( \, . \, ,b) $ is a measurable function on
$ \Omega $ and, for each $ \omega \in \Omega $,
$ K(\omega, \, . \, ) $ is a probability measure on $ \Upsilon $;
\[
b \mapsto \int_{\Omega} K(\omega,b) \, \mu(d\omega)
\]
is the probability distribution of the observable, resp., fuzzy random
variable $K$ in the state $ \mu \in {\cal M}(\Omega) $. Now, in the case
of a POVM $F$ on $ (M,\Upsilon) $, Eq.~(\ref{qc}) can be rewritten
according to
\begin{equation}\label{qc-povm}
{\rm tr} \, WF(b) = \int_{\ph} K(P,b) \, \mu(dP)
\end{equation}
where the Markov kernel $ K \! : \ph \times \Upsilon \rightarrow [0,1] $
is defined by $ K(P,b) := {\rm tr} \, PF(b) $. That is, every
quantum observable can be represented by a classical observable;
however, there are many more fuzzy random variables
$ K \! : \ph \times \Upsilon \rightarrow [0,1] $ than POVMs
$ F \! : \Upsilon \rightarrow \bsh $.

Summarizing, the statistical scheme of quantum mechanics can be
reformulated in classical terms by virtue of the Misra-Bugajski map.
This reformulation is complete in the sense that all quantum states
and quantum effects are represented as probability measures and
functions on the phase space $\ph$, respectively; however, not all
classically possible observables are quantum ones. Quantum mechanics
can thus be understood as a fuzzy probability theory on $\ph$ with a
selection rule for the observables; briefly, quantum mechanics is a
{\it reduced} fuzzy probability theory. Moreover, the projective
Hilbert space is a differentiable manifold carrying a natural
symplectic structure which allows one to reformulate quantum
dynamics in terms of Hamiltonian mechanics (cf.\
\cite{gue77,kib79,cir84,cir90,bro01,bje05}). Hence, quantum
mechanics can be interpreted to be a reduced classical statistical
mechanics on the phase space $\ph$.


As already observed by Bugajski in 1991, the classical embedding of
quantum mechanics induced by the Misra-Bugajski map contains all
ingredients of a hidden-variables, or ontological, model of quantum
mechanics. In fact, there is a phase space whose points may be taken
to play the role of {\em ontic} states describing the hypothetical
underlying reality of the quantum system. Next, there is the set of
probability measures $\mu$ over the phase space, which can be
interpreted as {\em epistemic} states describing the lack of
information about the actual ontic state in a preparation of the
system represented by $\mu$. Finally, there is the correspondence
(\ref{qc}) between quantum and classical expectation values which
determines the correspondences $\mu\mapsto W_\mu$ and $A\mapsto f_A$
between the quantum states and observables on the one hand and the
classical epistemic states and functions on phase space on the other
hand.

This ontological model is noncontextual with respect to measurements
since to every quantum effect probabilities are assigned that are
independent of the observables to which this effect may belong.
However, the model does display contextuality with respect to
preparations, in the sense defined by Spekkens \cite{spe05}: two
preparations that are statistically indistinguishable and hence
represented by one and the same density operator $W$ are generally
represented by different probability measures $\mu$ and $\mu'$ on
the phase space $\ph$ such that $W=W_\mu=W_{\mu'}$. This was
demonstrated in the proof of Theorem \ref{thm:mb}, part (b).

The function $P\mapsto K(P,b)$ appearing in (\ref{qc-povm}) can be
interpreted as the probability for the outcome of a measurement of
the observable $F$ to lie in the set $b$, given that the ontic state
of the system is $P$. This is to say that the present ontological
model constitutes a so-called stochastic or non-deterministic
hidden-variables model.

An ontological model of quantum mechanics can be said to ascribe
reality to the pure quantum states if any change in a pure state
must be associated with a corresponding change in the ontic state of
the system \cite{spe05}. The Misra-Bugajski map satisfies this
condition since the correspondence between pure quantum states and
point measures is given by a map $\delta_P\mapsto R\delta_P=P$.

In \cite{har04}, Hardy has given a proof of the fact that any
ontological model that reproduces the quantum mechanical
expectations must carry a large amount of ``quantum ontological
excess baggage"; more precisely, it is shown that even for a
finite-dimensional quantum system, any ontological model that
accounts for all quantum probabilities is based on a classical phase
space with infinitely many points, so that the epistemic states form
an infinite-dimensional simplex.

The requirements Hardy stipulates of an ontological model of quantum
mechanics are essentially those of our definition of a reduction map
$R$. If one accepts, in addition, the seemingly innocent
requirement that the adjoint map $R^*$ associates bounded quantum
observables with bounded measurable functions on phase space, then
Theorem \ref{thm10} asserts that, after removing redundant points
from the phase space, $R$ is related to the Misra-Bugajski map via
the map $i$ according to (\ref{R-mui}) and (\ref{r-rmb}), so that
essentially all ontological models  arise from some classical
reduction map as defined in the present paper. The uncountable
infinity of point measures in the set of epistemic states is now an
immediate consequence of Theorem \ref{thm10}.

It is evident that preparation contextuality is necessary for any
classical reduction map. As Examples \ref{ex2} and  \ref{ex3} show,
the correspondence $\delta_P\mapsto R\delta_P$ may be many-to-one,
and there may be point measures (hence ontic states) that are mapped
to mixed quantum states. The ontological model induced by the
Misra-Bugajski map is thus essentially distinguished (modulo
similarity) by a minimality or nonredundancy property in the sense
that a bijective correspondence is established between the pure
quantum states and the points of the associated classical phase
space. As Example 1 shows, this correspondence identifies Dirac
measures with pure quantum states up to a similarity transformation.

\section*{Acknowledgment}
This work was completed during W. S.'s visit at Perimeter Institute
(July-August 2007). Hospitality and support to both authors during
their visiting periods are gratefully acknowledged.

\end{document}